\documentclass[showpacs,showkeys,superscriptaddress,preprintnumbers,amsmath,amssymb,prl,floatfix,longbibliography]{revtex4}
\usepackage{graphicx}% Include figure files
\usepackage{dcolumn}% Align table columns on decimal point
\usepackage{bm}% bold math
\usepackage{amsmath}
\begin{document}

\title{Granular Impact: A Grain-scale Approach}

\author{Abram Clark}
\affiliation{Department of Physics \& Center for Non-linear and Complex Systems, Duke University, Science Drive, Durham NC 27708-0305, USA}
\author{Alec Petersen}
\affiliation{Department of Physics \& Center for Non-linear and Complex Systems, Duke University, Science Drive, Durham NC 27708-0305, USA}
\author{Lou Kondic}
\affiliation{Department of Mathematics, New Jersey Institute of
  Technology, Newark USA}
\author{Corey O'Hern}
 \affiliation{Department of Mechanical Engineering, Yale
  University, New Haven, USA}
  \author{R.P. Behringer}
  \affiliation{Department of Physics \& Center for Non-linear and Complex Systems, Duke University, Science Drive, Durham NC 27708-0305, USA}

\date{\today}

\begin{abstract}

This work summarizes a series of studies on two-dimensional granular
impact, where an intruding object strikes a granular material at high
speed. Many previous studies on granular impact have used a
macroscopic force law, which is dominated by an inertial drag term
proportional to the intruder velocity squared. The primary focus here
is on the microscopic force response of the granular material, and how
the grain-scale effects give rise to this inertial drag term. We show
that the inertial drag arises from intermittent collisions with
force-chain-like structures. We construct a simple collisional model
to explain the inertial drag, as well as off-axis instability and
rotations. Finally, we show how the granular response changes when the
intruder speed approaches $d/t_c$, leading to a failure of the
inertial drag description in this regime. Here, $d$ is the mean particle
diameter and $t_c$ the characteristic momentum-transfer time between
two grains.

%\note[LK]{LK to add something about simulations if needed. (This is supposed to be a book chapter, do we need abstract?}

\end{abstract}


\maketitle

\section{Introduction}

The physics of an intruding object impacting and penetrating a
granular material is relevant to many situations, including soil
penetration, meteor impacts, and ballistics. Related phenomena occur
in industrial settings, for instance in mixers or other machinery,
where a blade can impact a material. There is a large relevant
literature, of which we note a number of references, including the
references which they cite \cite{Allen1957,Forrestal1992,Tsimring2005,Katsuragi2007, Goldman2008, Goldman2010, Takehara2010, Clark2012, Clark2013, Clark2014, Ciamarra2004, Ambroso2005, deBruyn2004, Walsh2003, Nelson2008, Newhall2003, Seguin2009} %\cite{Poncelet1829, Allen1957, Forrestal1992, Tsimring2005, Katsuragi2007, Goldman2008, Goldman2010, Takehara2010, Clark2012, Clark2013, Clark2014, Ciamarra2004, Ambroso2005, Bruyn2004, Walsh2003, Nelson2008, Newhall2003, Campbell1986, Dexter2007}.

During an impact, a number of coupled processes are at work, and most
of these are only partially understood. At the largest scale, the
granular material exerts a force on the intruder, bringing it to
rest. The momentum and kinetic energy of the intruder are transferred
into the grains at the interface of the intruder and granular
material, and then carried deeper into the material. The time and space scales
that are relevant for these transfers must be determined to understand
the drag force on the intruder. 

A useful macroscopic starting point is a group of often-used empirical models for
the force experienced by the intruder. Here, we base our discussion on
a typical model (see also previous work
\cite{Allen1957, Tsimring2005, Katsuragi2007, Goldman2008, Goldman2010, Clark2012, Clark2013, Clark2014}) with the form:
\begin{equation}
F=m\ddot{z}=mg-f(z)-h(z)\dot{z}^2.
\label{eqn:forcelaw}
\end{equation}
$F$ is the force on the intruder, $z$ is the depth within the
material, where $z = 0$ corresponds to the point where the lower edge
of the intruder first touches the granular surface, $mg$ is the
gravitational force, $f(z)$ is a static term, $h(z)$ characterizes the
strength of the inertial ($\dot{z}^2$) term, and dots denote time
derivatives. Often, $h(z)$ is assumed to be constant, though we find
that it can have an initial transient. The static term, $f(z)$ is
often assumed linear in $z$. Although such models are empirical,
they typically capture the `slow' dynamics (i.e., neglecting short-time fluctuations) of intruder
trajectories. However, it is important to understand the grain scale
origins of these terms.

This equation is often dated back to Poncelet in the 1800s, and we will refer to it as the Poncelet model~\cite{Poncelet1829}. We understand it qualitatively in
terms of three forces which act on the intruder: gravity, a static
elastic or pressure-related force, and a collisional or momentum
transfer term, which should be proportional to the square of the
intruder speed, $\dot{z} = v$. The last point follows from the
collisional nature of the impact. During the majority of a typical
impact, the collisional (third) term dominates. But, when the intruder
comes to rest, the static term is important. This equation has been
tested in various settings, where it frequently provides a good
description of the motion of the
intruder~\cite{Allen1957,Katsuragi2007,Goldman2008,Goldman2010,Clark2012,Clark2013,Clark2014}.

However, Eq.~\ref{eqn:forcelaw} contains only the most basic physics,
and at this point, there is no direct connection to the particle
properties (such as the stiffness of the grains, their size relative
to the intruder, their shape) or to properties of the intruder (such
as its mass or shape). What are the grain-scale physical processes
which give rise to this equation? When and why might it break down? In
the discussion below, we show that the Poncelet equation works well to
describe the slowest time scales of the intruder dynamics in a series
of two-dimensional impact experiments into granular systems comprised
of photoelastic disks. We provide a deeper understanding of the origin
of the collisional term, proportional to $\dot{z}^2$, in terms of our
grain-scale observations.  Additionally, we look at how the
microscopic physics changes when the impact speed approaches the speed
at which forces propagate within the material, and we show how and why
the Poncelet model breaks down. Further insight into this regime of
fast impact is reached by discrete element simulations.

\section{Testing the form of the Poncelet equation}

\begin{figure} \centering  
\includegraphics[width=3in]{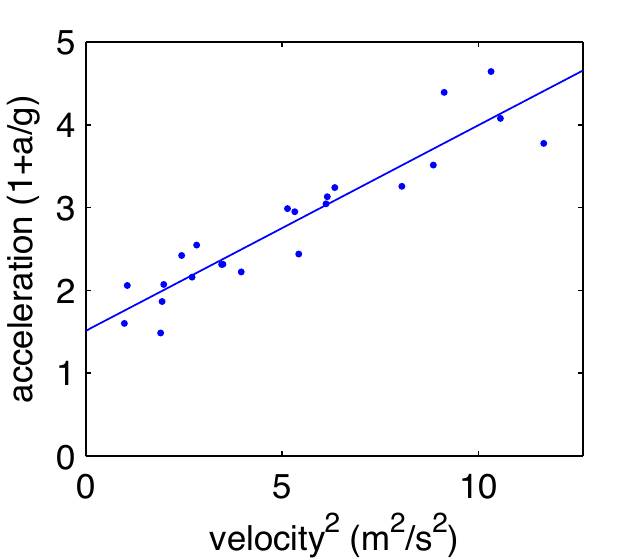}
\caption{Plot of $a$ versus $v^2$ for a circular
  intruder at a fixed depth, $z=\zeta$, where each data point
  represents one trajectory with varying initial velocities. The slope
  and offset of a linear fit to this data yield $f(\zeta)$ and
  $h(\zeta)$. However, large acceleration fluctuations make precise
  extraction of these values difficult without simplifying
  assumptions, e.g., $h(z)=b=$~constant, where $b$ is the average $h(z)$ for
  all depths. See Ref.~\cite{Clark2013} for more details.}
\label{fig:accvsvsqr}
\end{figure}

Testing the Poncelet equation requires data for the acceleration of
the intruder. Such data has been obtained and successfully fitted to a
Poncelet-like model in a few
cases~\cite{Goldman2008,Clark2012,Clark2013,Katsuragi2007}, almost
always in a regime where the intruder speed is much smaller than a
typical force propagation speed within the granular material. Here, we
consider data by Clark et al. \cite{Clark2012,Clark2013} for circular
intruders impacting beds of photoelastic particles from above. More
details are given in the experimental methods section, and for the
moment, we note that the results of this subsection pertain to initial
intruder speeds of up to $6$~m/s onto the hardest of the three types
of particles which we have
studied~\cite{Clark2012,Clark2013,Clark2014}. The impact speed in this
set of experiments is much less than the granular force transmission
speed (typically, in these materials, forces propagate at roughly
$v_f\approx 300$~m/s), and measurements of the acceleration of the
intruder are made on a time scale (after filtering) of $\sim
10^{-2}$~s. On this time scale, there are significant fluctuations of
the intruder acceleration, but it is still possible to obtain a `slow
time' measure of the intruder speed and acceleration as a function of
depth below the point of impact. The depth-dependent coefficients
$f(z)$ and $h(z)$ can be determined from a plot of the intruder
acceleration, $a$ plotted as a function of $v^2$. To obtain this plot,
we made multiple repetitions at different impact speeds in order to
map out the depth dependence of $f$ and $h$. Fig.~\ref{fig:accvsvsqr}
gives an example for a circular intruder.

As written, the Poncelet equation is a nonlinear second order equation
in $z(t)$. However, by recasting this equation using the identity
$m\ddot{z}=dK/dz$, familiar from basic mechanics, it becomes
(e.g. Eq.~\ref{eqn:KEmodel}) a first order linear equation for $K(z)$, the
kinetic energy, as a function of depth, with non-constant
coefficients. it is then possible to write a formally exact solution
for $K(z)$.

\begin{equation}
\frac{dK}{dz} = mg-f(z)-\frac{2h(z)}{m}K.
\label{eqn:KEmodel}
\end{equation}

Once $K(z)$ is determined by integrating Eq.~\ref{eqn:KEmodel}, $z(t)$
follows by writing $\dot{z} = (2K(z)/m)^{1/2}$, integrating $t(z)
=\int_0^z dz (2K(z)/m)^{-1/2}$, and inverting. If the forms of $f(z)$
and $h(z)$ are simple, the calculation can by done explicitly. For
example, using the commonly assumed forms $f(z)=f_0+kz$ and $h(z)=b$,
we obtain $K(z)$ as
\begin{equation}
K(z) = (K_0 -c_1)\exp(-c_2 z) + c_1 - c_3 z.
\label{eqn:k-z-simp}
\end{equation} 
Here, the constants are $c_1 = [(mg-f_0)c_2 + k]/{c_2}^2$, $c_2 =
2b/m$, and $c_3 = k/c_2 = km/(2b)$.

Even without integrating, it is possible to find the stopping distance
by setting $K(z_{stop}) = 0$, which yields the stopping depth as a
function of impact energy, $K_0$. For the common case described by
Eq.~\ref{eqn:k-z-simp}, the stopping depth, $z_{stop}$, varies as the
log of $K_0$:
\begin{equation}
z_{stop}=\frac{m}{2b}\log\left[\frac{\frac{2b}{m} K_0+\left(f_0+\frac{km}{2b}\right)-mg}{\left(f_0+kz_{stop}+\frac{km}{2b}\right)-mg}\right].
\label{eqn:z-stop-simp}
\end{equation}
If $f(z)$ as roughly constant, $f(z)=f_0$ and $k=0$, then the form for
the stopping distance is simpler. In this case, $z_{stop}$ still
increases logarithmically with $K_0$, as in \cite{Tsimring2005,
  Goldman2008, Forrestal1992}.
\begin{equation}
z_{stop}=\frac{m}{2b}\log\left[1+\frac{2b}{m}\left(\frac{K_0}{f_0-mg}\right)\right]\ .
\label{eqn:highKEsol}
\end{equation}
Such an approximation is increasingly relevant as the intruder speed
increases, and the $\dot{z}^2$ term dominates.

\begin{figure} \centering
\hspace{0.5in}\raggedright (a) \hspace{3in} (b) \\ \centering
\includegraphics[width=3in]{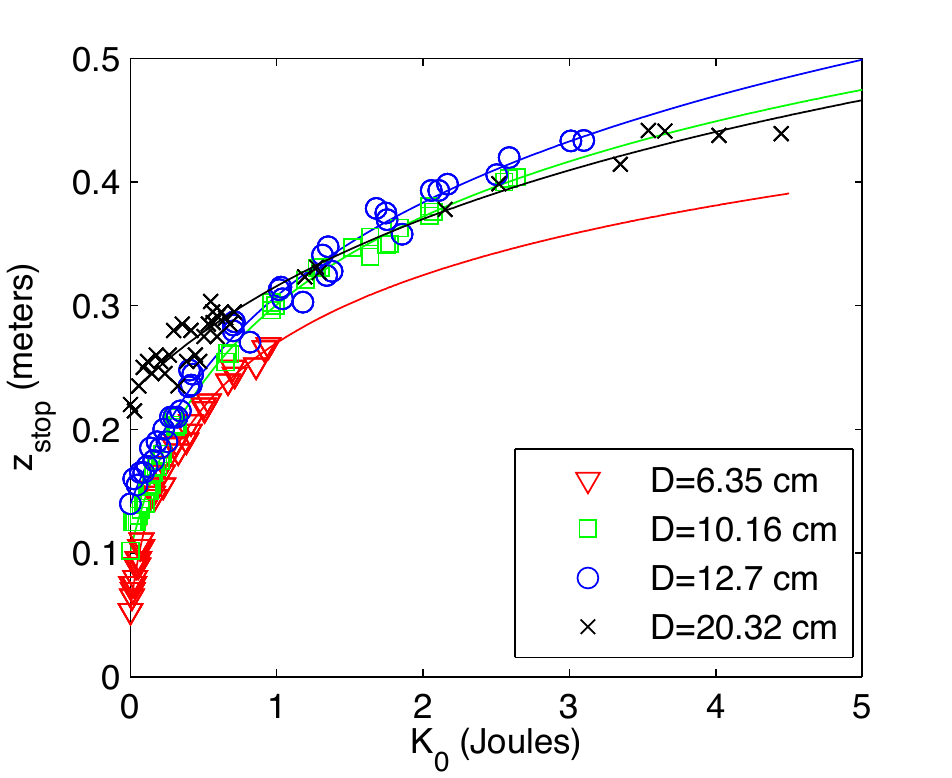} 
\includegraphics[width=3in]{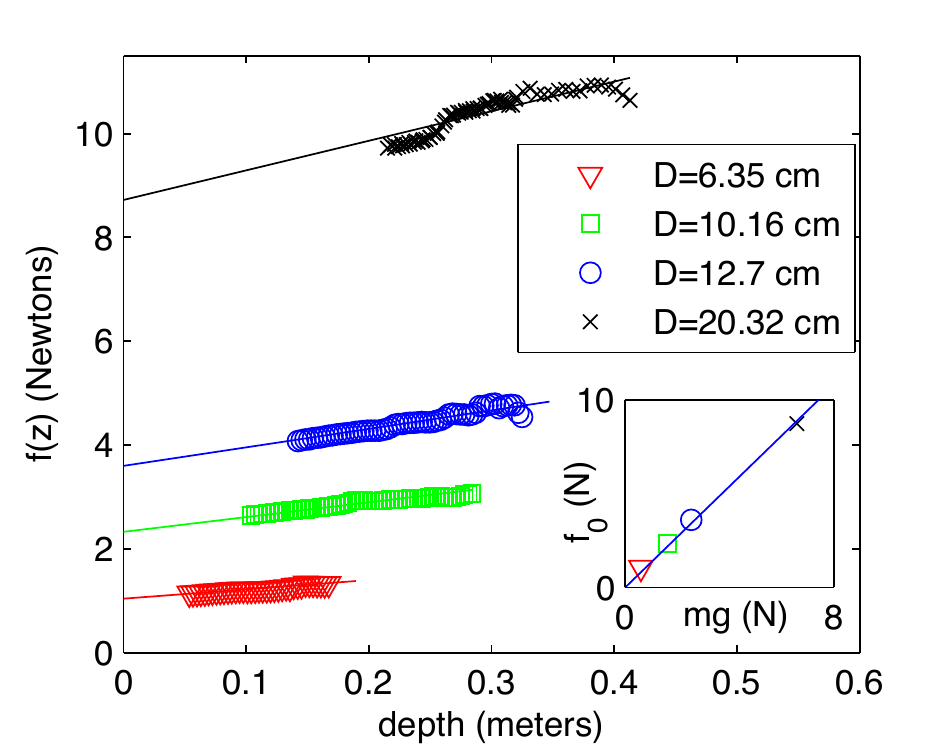} \\
%\onefigure[scale=0.7,trim=0mm 0mm 0mm 0mm]{fzcircles.eps}
\caption{
(a) Plot of $z_{stop}$ vs. $K_0$, with fits of the form
  $a\log (bK_0+1)+c$. (b) Plot of $f(z)$ for circular
  intruders. Linear fits are $f_0+kz$, where the slope, $k$,
  corresponds to hydrostatic pressure. Inset shows plot of $f_0$
  vs. $mg$, with a linear fit through the origin, with slope 1.35.}
\label{fig:fzplots}
\end{figure}

\begin{figure} \centering
\hspace{0.5in}\raggedright (a) \hspace{3in} (b) \\ \centering
\includegraphics[width=3in]{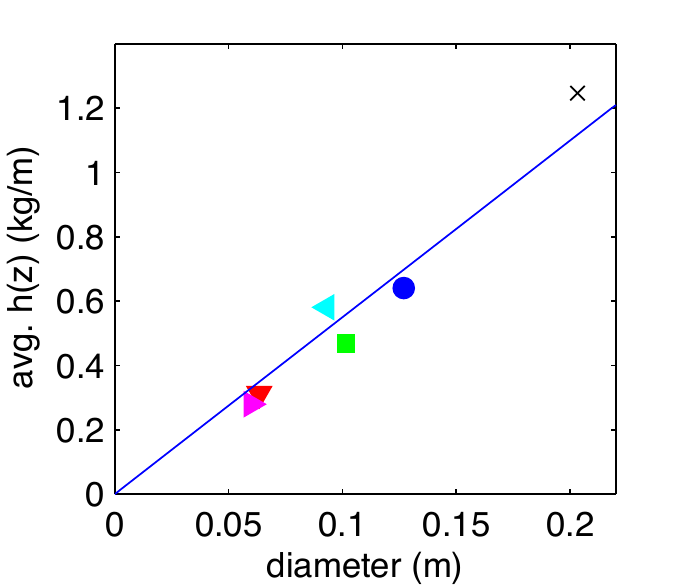}
\includegraphics[width=3in]{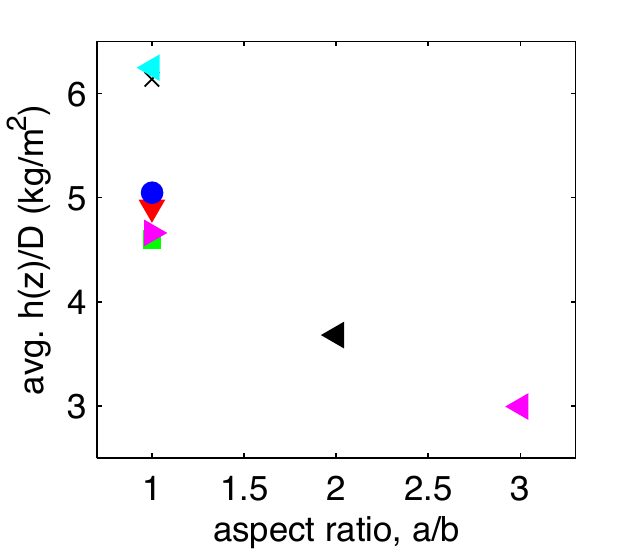}
%\onefigure[scale=0.78,trim=0mm 0mm 0mm 0mm]{hzvsaspectratio.eps}
\caption{ {(a) Plot of the average $h(z)$ for circular-nosed intruders
    versus $D$, which shows that $h(z)\sim 5.5D$. (b) Plot of the
    average $h(z)/D$ versus the intruder aspect ratio, $a/b$, which
    shows a substantial decrease $h$ as the
    intruder nose is elongated.}}
\label{fig:hzvsshape}
\end{figure}

Recasting the Poncelet equation in terms of kinetic energy is also
advantageous in fitting experimental data to the Poncelet equation,
since only velocity data is needed, and not acceleration data.  For
further discussion on this, see~\cite{Clark2013}. In
Fig.~\ref{fig:fzplots}(b), and Fig.~\ref{fig:hzvsshape}, we show data
for $h$ and $f(z)$, fitted using the kinetic energy approach. The
static term $f(z)$ is a linear function of $z$ to a good
approximation. The drag coefficient, $h(z)$, is nearly constant after an initial
transient near the free granular surface, and we show the average drag coefficient, $h_0$, as a function of intruder size and shape in and Fig.~\ref{fig:hzvsshape}.


\section{A grain-scale force picture}

\begin{figure*}[th!] 
%\centering
%\includegraphics[clip,trim=8mm 14mm 8mm 10mm,width=\columnwidth]{spacetimewdecay.eps} 
\raggedright (a)\\ \includegraphics[clip,trim=18mm 3mm 15mm 3mm,width=\textwidth]{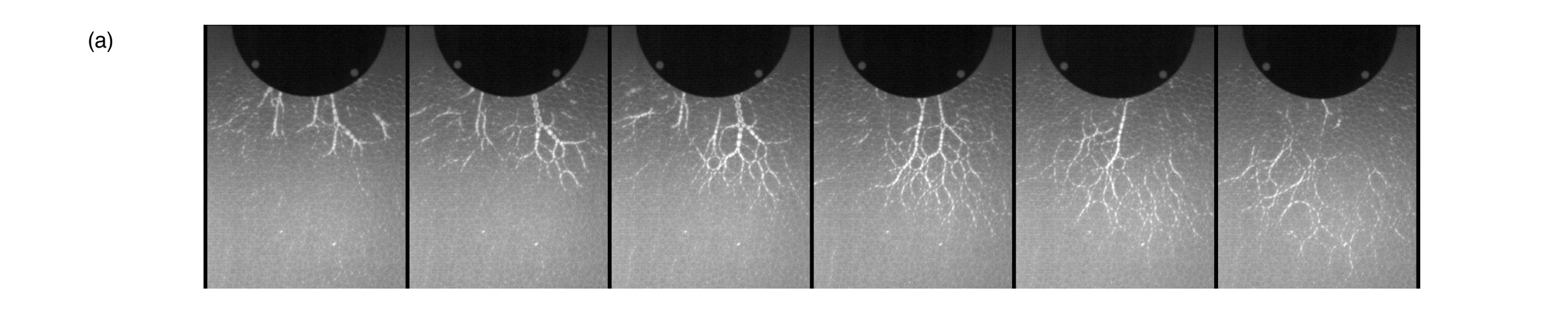}
\raggedright (b)\\ \includegraphics[clip,trim=18mm 5mm 15mm 0mm,width=\textwidth]{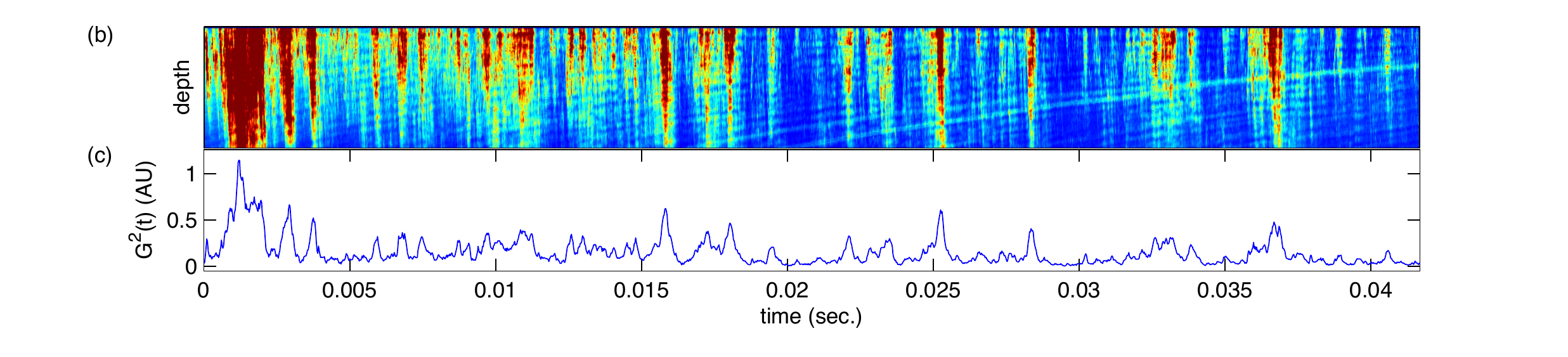}
 \caption{(a) Six frames (starting at 2.75 ms after impact and spanning 475 $\mu$s) showing how forces propagate from the intruder into the granular material. (b) 
     A space-time plot for the response under the
     bottom half of the intruder over time (x-axis is time, and
     y-axis is radial distance from the intruder, spanning roughly 10 particle diameters). The slope of these lines gives a force propagation speed  of roughly 300~m/s. (c)
     The sum of the response in the space-time plot above, with background lighting subtracted. This time series yields a measurement of instantaneous force on the intruder, where the range shown above (0 to 1.1 AU)
     maps to an acceleration range of 0 to 27 g.}
 \label{fig:frames}
\end{figure*}

The collisional term, involving the coefficient $h(z)$ is the dominant
drag term for much of the duration of a collision. What are the
grain-scale physical processes which give rise to this term?
Figure~\ref{fig:frames} shows a series of photoelastic images, taken
from a high-speed movie of an impact, where bright particles are
experiencing a strong force. These images suggest a qualitative
picture for the case we are studying: in the regime where the Poncelet
equation is valid and the intruder speed is well below the granular
force propagation speed, the intruder excites forces along filimentary
structures, that are similar in nature to the force chains of static
or quasi-static deformations of granular materials
\cite{Radjai1996,Liu1992,Howell1999,Baxter1991,Dantu1968,Wakabayashi1950,Drescher1972}. However, unlike in the static case, forces propagate
dynamically along the filimentary force network. The network structure
evolves in time as the intruder plows through the material, but it is
relatively long-lived compared to the time for a signal to propagate
along a segment of the network.

\begin{figure} \centering
	\includegraphics[clip,trim=7mm 0mm 10mm 7mm,width=0.6\columnwidth]{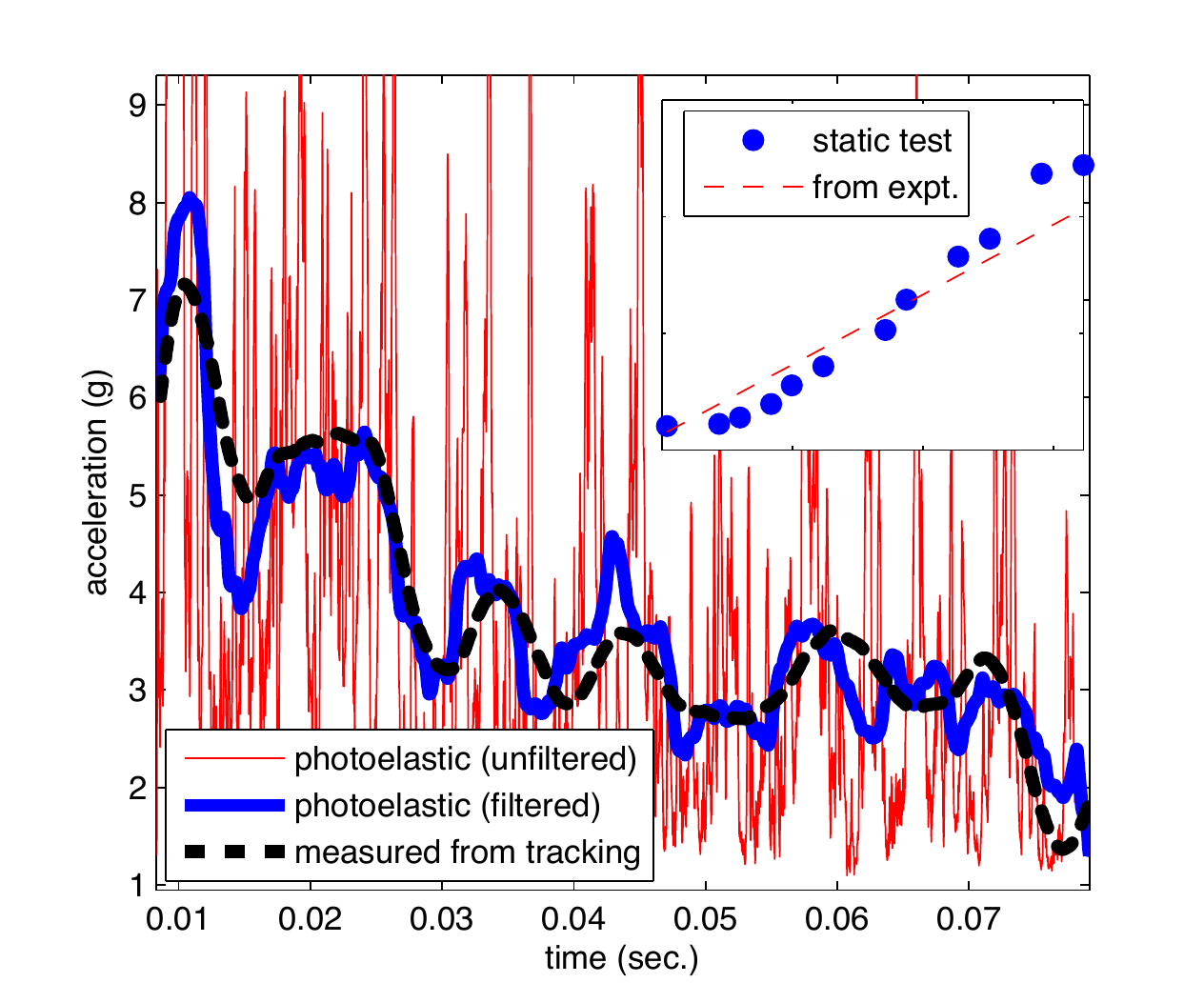}
	\caption{Plot showing good correlation between acceleration fluctuations and
          photoelastic response. The acceleration measurement (black,
          dashed line) requires time-filtering which limits the high
          frequency resolution. We average the photoelastic response from Fig.~\ref{fig:frames}(b)
          such that the frequency resolution is the same as for the
          acceleration (blue line). With a linear rescaling of the photoelastic
          measurement, we observe extremely close agreement, both for the mean and fluctuations. The thin
          red line shows the calibrated photoelastic force measurement
          without time-filtering, which shows very large fluctuations
          at short time scales.}
	\label{fig:accphoto}
\end{figure}

The forces carried by the force network fluctuate at high frequencies,
as suggested by the images shown in Fig.~\ref{fig:frames}. An
immediate question is, do the fluctuations in the photoelastic signal
really correspond to fluctuations that are transmitted to the
intruder? We can provide at least a partial answer to that question by
matching the photoelastic signal to the measured acceleration of the
intruder. This acceleration is determined by tracking the center of
mass of the intruder, and then differentiating twice. Numerical
differentiation accentuates the tracking errors, and it is necessary
to low-pass filter the computed intruder acceleration with a cutoff
which we set to 133 Hz. The photoelastic response is obtained by
summing the pixel values in the photoelastic images in a region
beneath the intruder (after correcting for inhomogeneity in the
background lighting).  We then filter the photoelastic response in a
manner similar to the acceleration, and find that there is a very good
agreement between the filtered photoelastic force signal and the
intruder acceleration, as in Fig.~\ref{fig:accphoto}. This suggests
that the photoelastic fluctuations (red curve in
Fig.~\ref{fig:accphoto}) correspond to fast, physical, force
fluctuations acting on the intruder.

The scenario that we suggest is that as the intruder plows through the
granular material, it collides with a latent network, which then
responds by carrying momentum and energy into the material. By latent,
we mean that a subset of grains are initially positioned so that when
the intruder strikes, they can respond as a `stiff' network, even
though they were not necessarily strongly compressed before impact. A
slightly different scenario is to think of the parts of the network as
relatively independent `clusters' that collide with the intruder. A
cluster here is highly anisotropic--i.e. it is a quasi-linear
collection of particles.

\begin{figure} \centering
 \includegraphics[clip,trim=5mm 0mm 5mm 0mm,width=0.6\columnwidth]{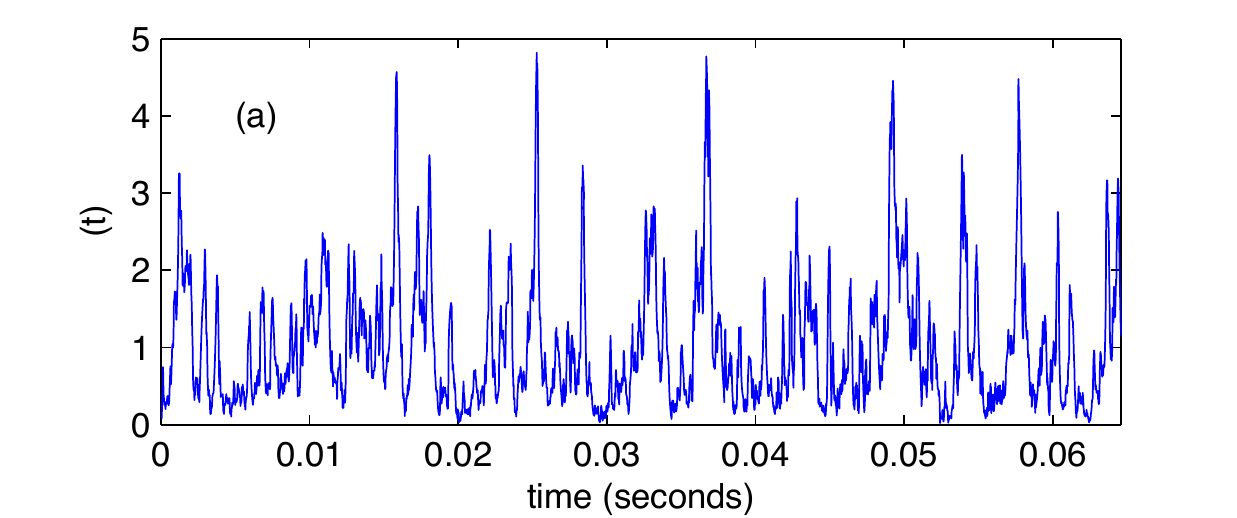}
 \includegraphics[clip,trim=5mm 0mm 5mm 0mm,width=0.6\columnwidth]{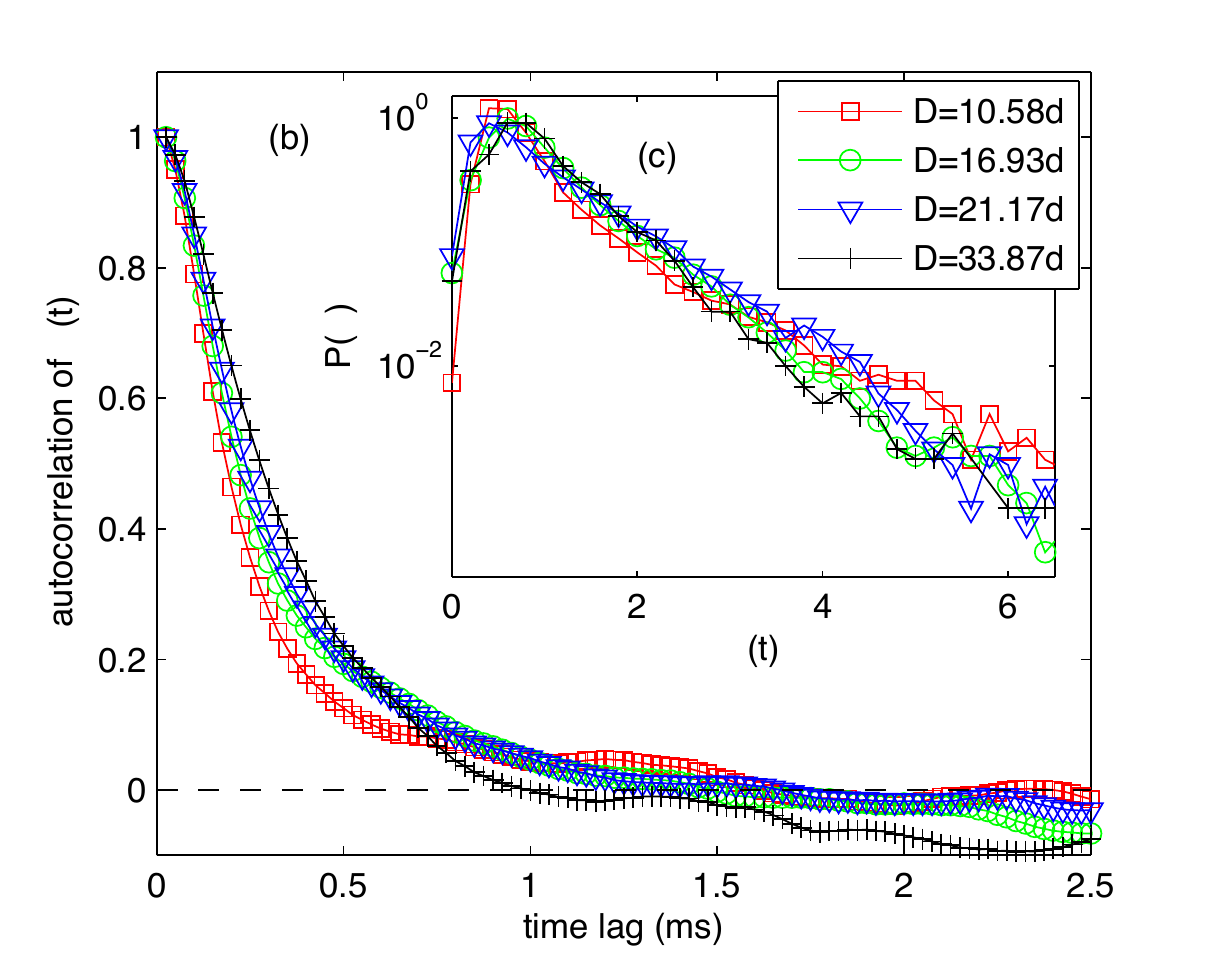}
 \caption{(a) Plot of $\eta(t)$
     for a single experiment (typical of all experiments), where $\eta(t)\sim F(t)/F_{slow}(t)$, as
     discussed in the text. (b) The autocorrelation
     and (c) the PDF (semi-log scale) of the combined fluctuating signals
     for all experiments for each intruder ($\sim$20 experiments per intruder). We observe $P(\eta)\sim
     \exp(-\eta)$ with an autocorrelation length of roughly 1
     ms, which gives a typical event time, in agreement with Fig.~\ref{fig:frames}.}
 \label{fig:flucts}
\end{figure}

\begin{figure} \centering
\includegraphics[width=0.6\columnwidth]{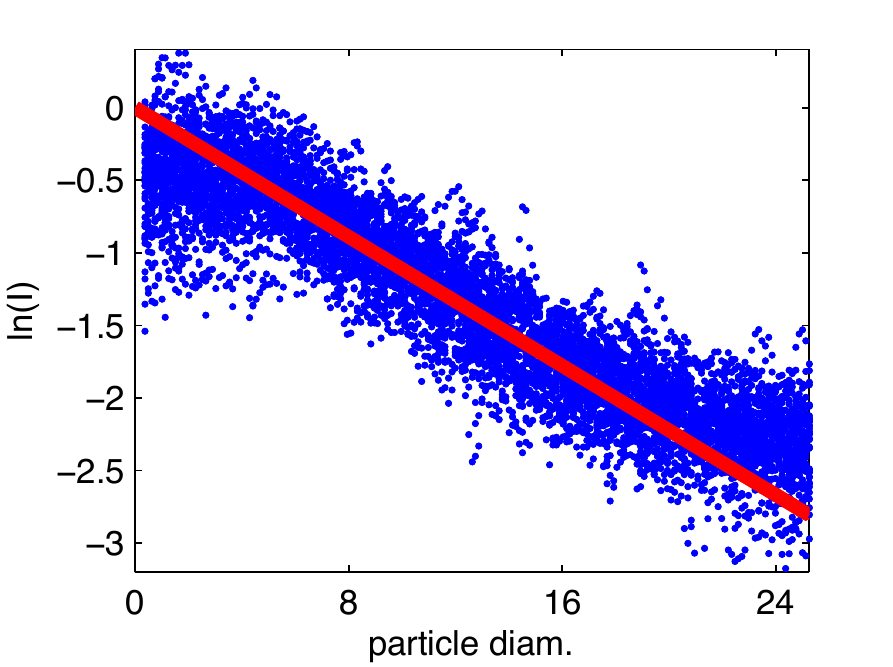}
\caption{Using a thin angular slice (width of $\pi/8$ radians) centered directly
    beneath the intruder for the 10 fastest experiments for one intruder size, we identify the 4 largest acoustic
    pulses in each run. We then plot the natural logarithm of acoustic
    intensity versus depth for each pulse, normalized by the
    total intensity in the pulse. The red line shows a fit of
    $\exp(-r/L)$, where $L$ is the decay length. For this intruder, we
    observe a decay length of about 10 particle diameters.}
\label{fig:decay}
\end{figure}

We further pursue the nature of the force fluctuations and their
connection to a collisional damping term by normalizing the force
signal by the `slow time' drag force, $F_{slow}$ (i.e., a low-order
polynomial fit to $F(t)$) signal, to produce data such as
Fig.~\ref{fig:flucts}(a). In Fig.~\ref{fig:flucts}(b) we show the time
correlation function of the fluctuating term, $\eta = F(t)/F_{slow}$,
which decays with a characteristic time of $\tau_d \simeq 1$ ms. We
also determine the penetration range of the strong force network
vs. depth below the intruder, as in Fig.~\ref{fig:decay}. These data
are consistent with an exponential decay of the strong response
vs. distance below the intruder, with a decay length of $L_d \simeq 10
d$, where $d$ is a mean grain diameter of roughly $5$~mm. Note that
the ratio of decay length to the decay time is $L_d/\tau_d \simeq 50$
m/s, which is faster than the intruder speed (roughly $5$ m/s) and
slower than the granular sound speed (roughly $300$ m/s) for the
material shown here.

\section{Collisional Model}

Our observations thus far suggest that momentum is transferred to the
granular material through intermittent collisions with a force
network, sending momentum and energy away in the form of pulses. To
capture the basic physics of this process, we will present a model
that assumes collisions with grain `clusters', which are understood to
be collections of particles forming part of the strong force
network. These collisions occur along the lower boundary of the
intruder at random locations (during the course of a complete
impact). The model contains several assumptions:
\begin{itemize}
\item
Momentum transfer from intruder to granular material takes place by
discrete collisions with clusters (as discussed above) with a typical
mass which is large compared to that of a grain, but small compared to
the system size.
\item
Collisions occur, and transfer momentum in the direction normal to the
intruder boundary. This assumption is generally consistent with visual
inspection of the force chain directions.
\item
Collisions occur inelastically with a typical restitution coefficient, $e$.
\item
The rate of collisions is set by the instantaneous intruder speed, $v$.
\end{itemize}

\begin{figure} \centering
\includegraphics[width=0.5\columnwidth]{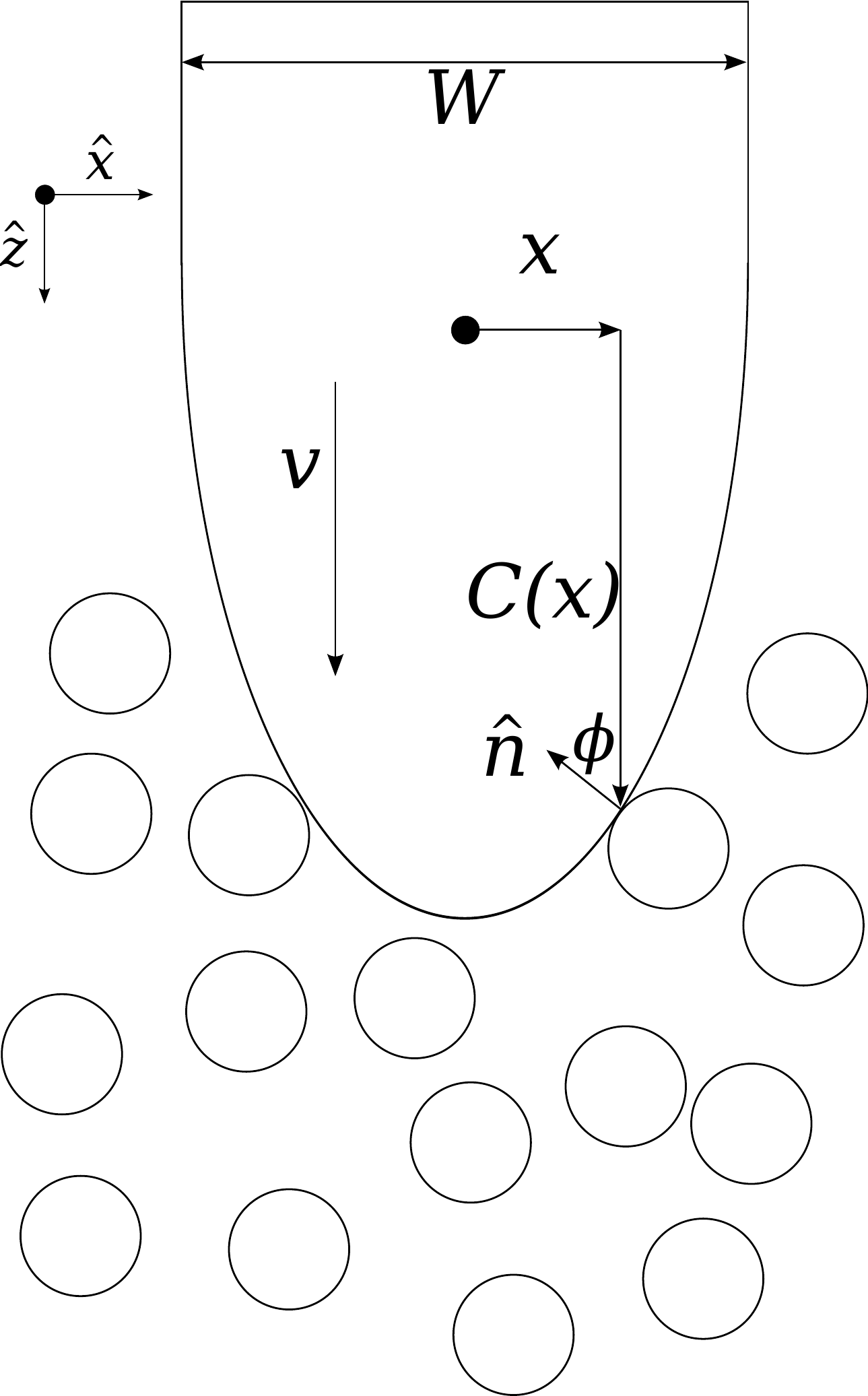}
\caption{Sketch of the collisional model, where an intruder of width
  $W$ randomly collides with grain clusters (represented by the open
  circles) as it moves at velocity $v$. The collisions occur along the
  `nose', i.e. the leading edge of the intruder, at positions
  $\vec{r}=x\hat{x}+C(x)\hat{z}$ measured from the center of mass.
  These collisions involve momentum transfer normally into the
  intruder, along normal vector $\hat{n}$.}
\label{fig:cartoon}
\end{figure}

These assumptions lead naturally to collisional damping proportional
to $v^2$. Additionally, it is possible to test this picture by considering how
the shape of the intruder affects the dynamics. If, as in
Fig.~\ref{fig:cartoon}, $\hat{n}$ is the local normal to the lower
boundary of the intruder, then the force acting on the intruder from a
collision is

\begin{equation}
\vec{f}=\frac{\Delta\vec{p}}{\Delta t}=\frac{(1+e) v^2\cos^2\phi}{\gamma d}\left(\frac{m_cm}{m_c+m}\right)\hat{n}.
\label{eqn:collforce}
\end{equation}

The $z$-component of the total force acting on the intruder is then
\begin{eqnarray}
F_z&=&\int d\vec{F}\cdot\hat{z}\nonumber \\
&=&\frac{(1+e)\beta m_cm}{\gamma d^2(m_c+m)}\left[\int_{-W/2}^{W/2}dx(1+C'^2)^{-1}\right]v^2\nonumber \\
&=& B_0\cdot I[C(x)]\cdot v^2.
	\label{eqn:zforcecoll}
\end{eqnarray}
\noindent
The total torque, $\vec{\tau}$, about the center of mass of the
intruder is given by integrating $\vec{r} \times \vec{f}$ over all
collisions at the intruder surface, similarly to
Eq.~\ref{eqn:zforcecoll}:
\begin{eqnarray}
\vec{\tau} &=& \int \vec{r}\times\vec{f}\,\,\frac{\beta}{d}dl \nonumber \\
	&=& B_0v^2\int\vec{r}\times \hat{n}\cos^2\!\phi\, dl
\label{eqn:torque-tot}
\end{eqnarray}
Here, $\vec{r}=x\hat{x}+C(x,\theta)\hat{z}$,
$\hat{n}=-\sin\phi\hat{x}-\cos\phi\hat{z}$, and
$\sin\phi=-C'(1+C'^2)^{-1/2}$, with $dl=(1+C'^2)^{1/2}dx$ and
$\cos\phi=(1+C'^2)^{-1/2}$, as before (see the sketch in
Fig.~\ref{fig:cartoon}). This
yields:
\begin{eqnarray}
\vec{\tau} &=& B_0v^2 \int dx \left(\frac{CC'}{1+C'^2}+\frac{x}{1+C'^2}\right)\hat{y}\nonumber \\
	&=& B_0 J[C(x,\theta)]v^2 \hat{y}.
\label{eqn:torque-tot2}
\end{eqnarray}

The mean force and torque expressions contain boundary integrals which
explicitly depend on boundary shape. The constant
$B_0=(1+e)\beta m_cm/(\gamma d^2(m_c+m)$ contains various
system parameters that we assume to be nominally the same for all intruders
striking a particular material. The size and shape effects are
contained in $I[C(x)]$, defined as:
\begin{equation}
I[C(x)]\equiv\int_{-W/2}^{W/2}dx(1+C'^2)^{-1},
\label{eqn:I}
\end{equation}
and in the corresponding functional $J$ for the torque. By
construction, the force on the intruder is proportional to
$v^2$. There is, in principle, only one free coefficient, $B_0$, which
contains all the unknown material-dependent properties, such as the
effective coefficient of restitution and the mass of a cluster. 
However, we find that if the intruder has a sharp tip, then this single-coefficient
approach must be replaced by two terms: one for the tip, and one for
the rest of the intruder surface.
\begin{figure*}[th!] 
\centering
 \includegraphics[clip,trim=35mm 0mm 35mm 3mm,width=\textwidth]{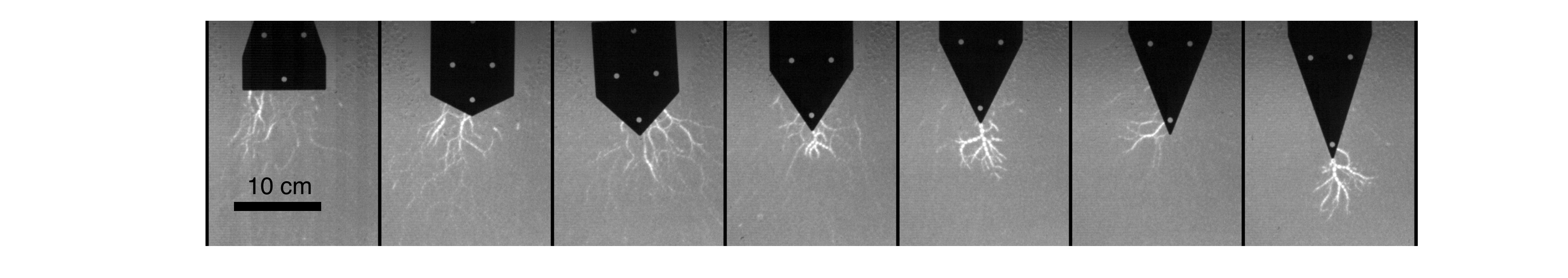}
 \caption{Still frames showing the seven triangular-nosed
   intruders with slope $s=0$
   to $s=3$ (increments of 0.5), from left to right. These images are chosen during photoelastic
   events. Note that the force chains are
   roughly normal to the
   intruder-granular interface. Additionally, these images illustrate that, 
   when $s$ is large,
   significant forces occur at the intruder tip.}
 \label{fig:triangles-photoel}
\end{figure*}

To test this model, we determine the coefficient $h(z)$ for intruders
of different shapes: circular, ogive, and triangular. By ogive, we
mean any shape consisting of a horizontally symmetric convex curved
leading edge of the intruder, such as half of an ellipse, joined above
to a straight rectangular tail, as in the sketch of
Fig.~\ref{fig:cartoon}.

%We show the various shapes which we
%characterized in Fig.~\ref{fig:trajectories}. 
%\begin{figure} \centering
%\includegraphics[width=3in]{depthvelacc.eps}
%%\onefigure[scale=0.85,trim = 0mm 25mm 0mm 28mm]{depthvelacc.eps}
%\includegraphics[width=3in]{intruders2.eps}
%%\onefigure[scale=.35,trim=0mm 0mm 0mm 0mm]{intruders2.eps}
%\caption{(TOP) Single trajectories (depth, velocity, and acceleration) for the three smallest circular intruders
%  with similar initial impact velocities, where $t=0$ is the first contact with the granular surface, measured
%  from photoelastic response. (BOTTOM) Shapes of all intruders, drawn to scale, as described in the text.}
%\label{fig:trajectories}
%\end{figure} 

\begin{figure} \centering
\hspace{0.5in}\raggedright (a) \hspace{2.3in} (b) \\ \centering
\includegraphics[clip,trim=10mm 18mm 10mm 10mm,width=2.3in]{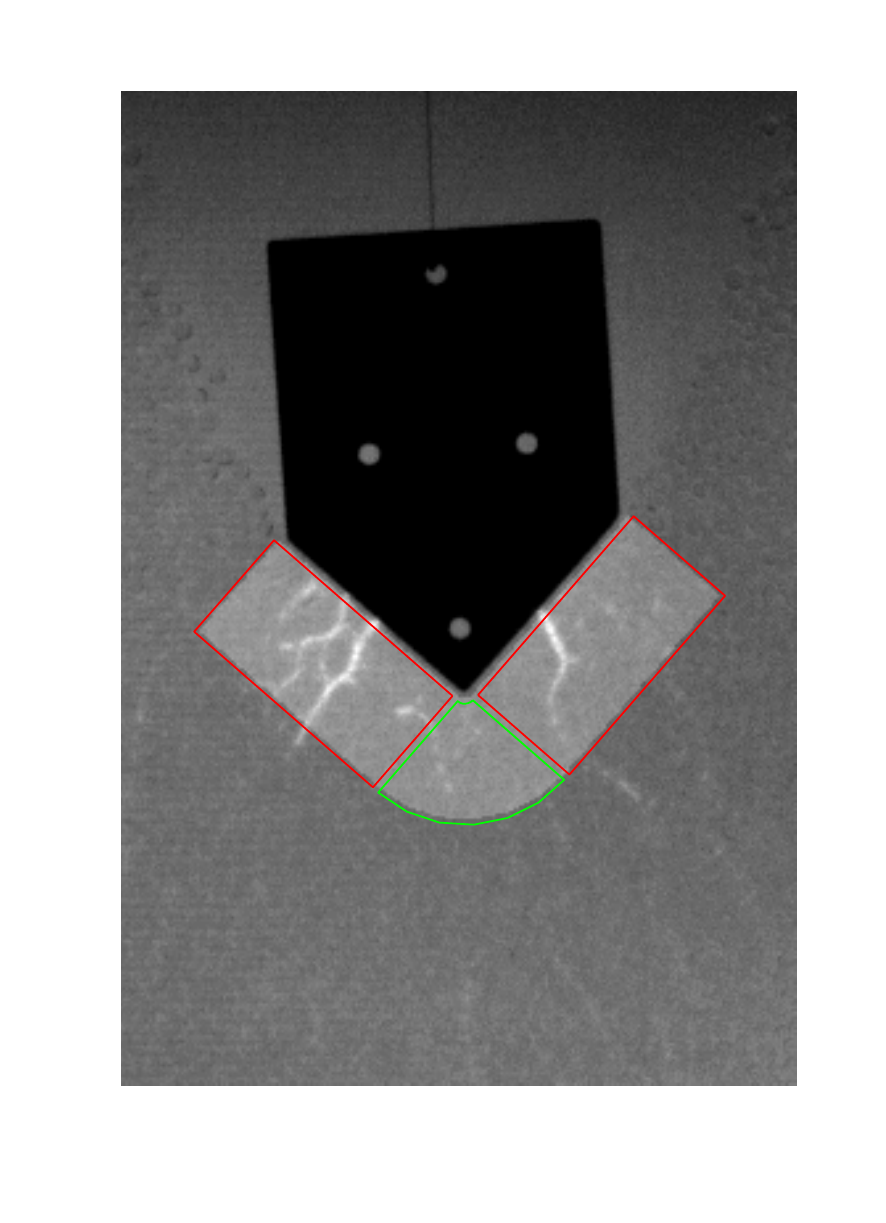}
\includegraphics[clip,trim=0mm 0mm 0mm 0mm,width=4in]{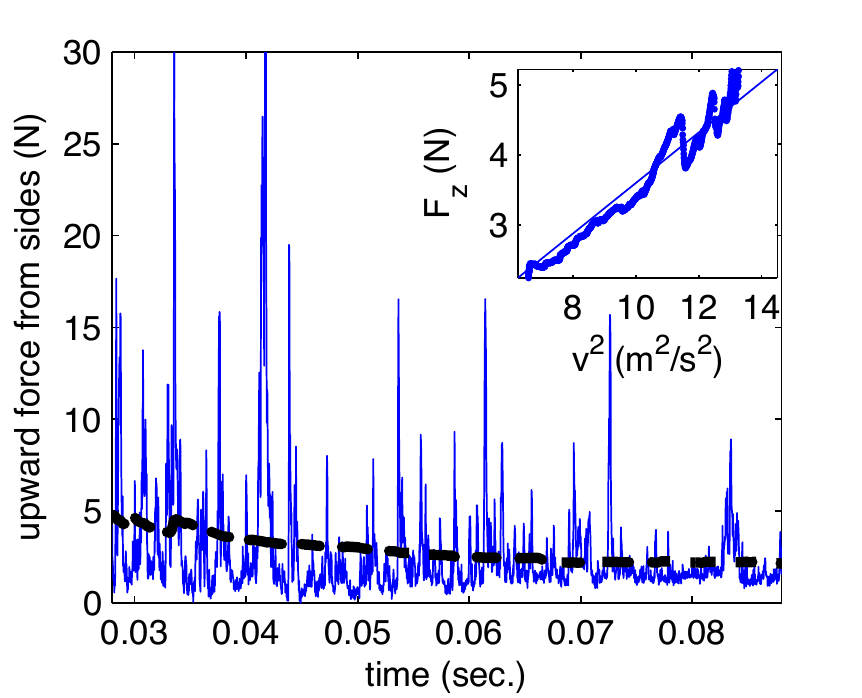}
\caption{(a) A photoelastic image showing the
  different regions used to measure the force contributions from the
  sides (outlined in red) and from the tip (outlined in green). (b) Plot of calibrated
  photoelastic force on the sides of the $s=1$ intruder, for a single
  trajectory with $v_0\approx 3.8$~m/s, both at each frame (thin, blue
  line) and after a low-pass filter (thick, black, dashed line). Inset
  shows the low-pass filtered force signal determined photoelastically
  vs. $v^2$, with a fit line through the origin with
  slope 0.36, which is the effective drag coefficient contribution
  from the sides of the triangular intruder in this case.}
\label{fig:photodrag}
\end{figure}

Triangular intruders are particularly simple because the shape
integrals involve a constant slope, $s$ for the leading edge. In
Fig.~\ref{fig:triangles-photoel} we show typical photoelastic images
from high speed videos of triangular intruder impacts. Here, $s$ is
the magnitude of the slope of the lower parts of the intruder relative
to the horizontal direction. For the more pointed triangles, it is
clear that the tip is more effective than the rest triangle boundary
at generating a strong force response. When analyzing triangles, we
consider the photoelastic response from the tip region separately from
the response on the edges, as in Fig.~\ref{fig:photodrag}, top.

\begin{figure} \centering
\hspace{0.5in}\raggedright (a) \hspace{3in} (b) \\ \centering
\includegraphics[clip,trim=0mm 0mm 0mm 0mm,width=3in]{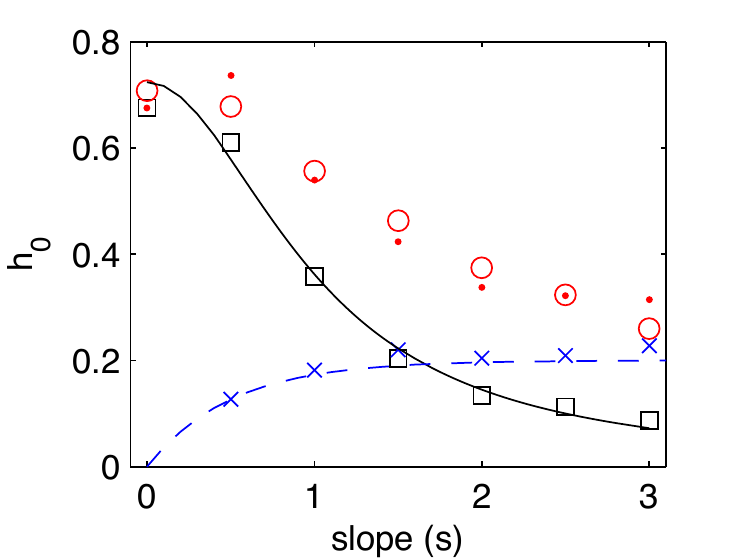}
\includegraphics[width=3in]{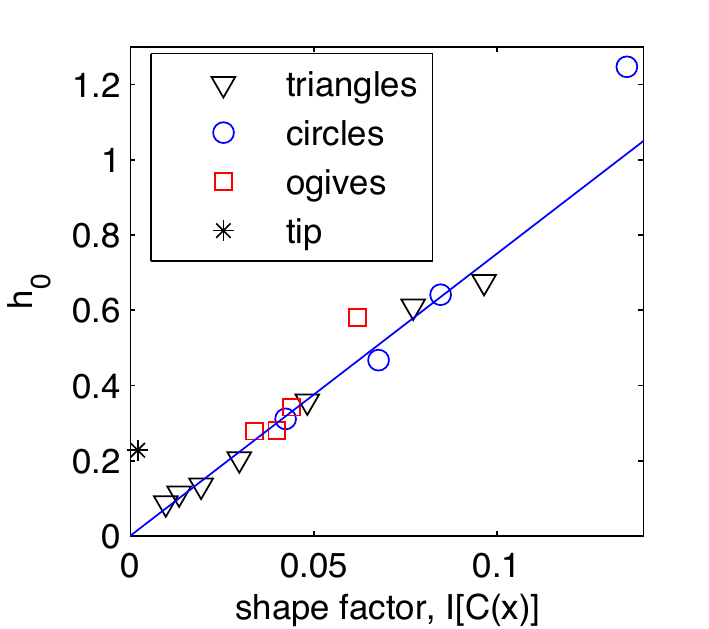}
\caption{(a) Plot of $h_0$ measured from the
  intruder acceleration and force measurements inferred from the
  photoelastic response. The measurements from intruder acceleration
  (open, red circles) and from the photoelastic signal (red
  dots) show good agreement. The contribution from the tip (blue crosses), measured from
  the photoelastic signal, stays relatively constant for $s>1$ (fit line is $0.2(1-e^{-2s})$,
  an approximate fit assigned by eye). Open,
  black squares show the contribution from the sides, which matches
  well to the model (solid black line),
  $I(s)=W(1+s^2)^{-1}$, as in
  Eq.~\eqref{eqn:shape-eff-triang}. (b) Plot of $h_0$ versus
  $I[C(x)]$ for all intruders. The solid line shows a linear fit
  through the origin with slope
  $\frac{h_0}{I[C(x)]}=B_0\approx 7.6$. The $h_0$ value for triangular
  noses is the photoelastic measurement from the sides (excluding the
  tip), while the asymptotic value (approximately 0.2 N) for the tip
  measurement is shown separately (black asterisk). See Refs.~\cite{Clark2013,Clark2014} for further discussion.}
\label{fig:hvsI}
\end{figure}

Since $h(z)$ is nearly constant, after small initial transients, we
show results in Fig.~\ref{fig:hvsI} for $h_0$, the constant part of
$h$, vs. the triangle slope, $s$. In this figure, we separate the
contributions from the tip and the flat surfaces. Note that for the
former, this effect is zero for $s = 0$, and then asymptotes to a
constant value as $s$ grows.

In Fig.~\ref{fig:hvsI}(b), we show data for all shapes, plotted as
$h_0$ vs. the shape factor from the collision model, $I[C(x)]$. The
data are in good agreement with the prediction from the model. The
proportionality constant, $B_0 \simeq 7.62$, then encompasses all the
unknown particle properties.

\begin{figure} \centering
\includegraphics[clip, trim=2mm 0mm 2mm 0mm, width=3in]{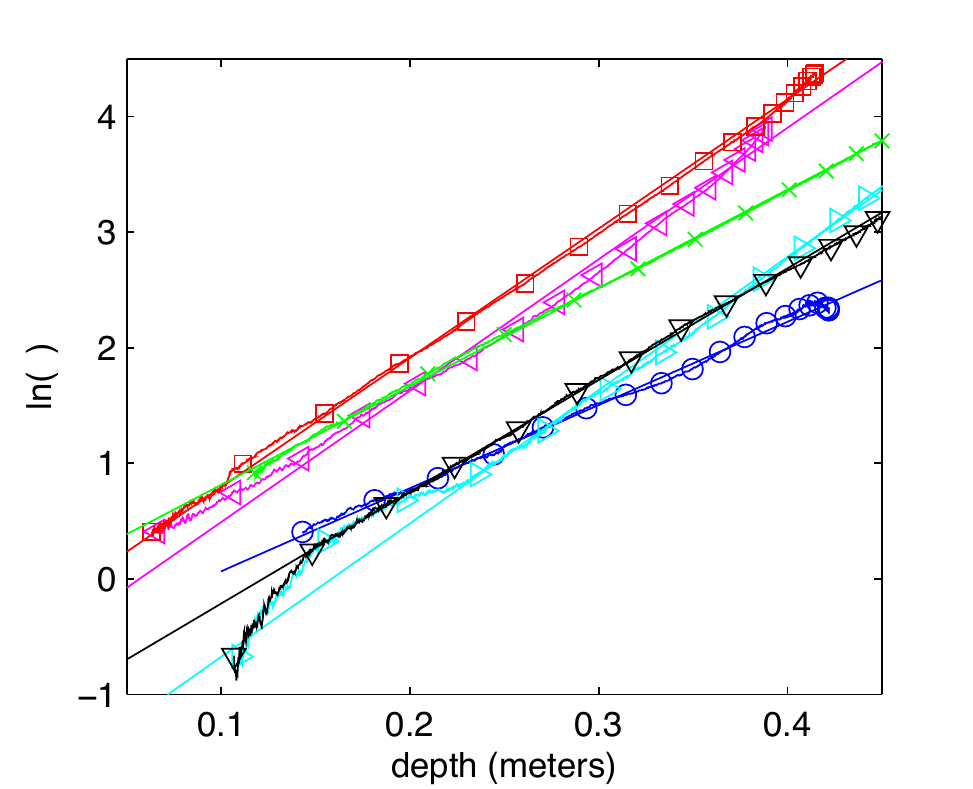}
\caption{Plot of the natural log of the angular
  deviation versus depth from single trajectories, where all angular
  deviations are considered positive. The straight lines suggest
  exponential growth for $\theta$ vs. depth; the slope on the semi-log plot
  corresponds to the exponential growth rate, $\lambda_+$, as
  discussed in the text.}
\label{fig:rotations}
\end{figure}

\begin{figure} \centering
\includegraphics[clip, trim=0mm 0mm 0.5mm 3mm, width=3in]{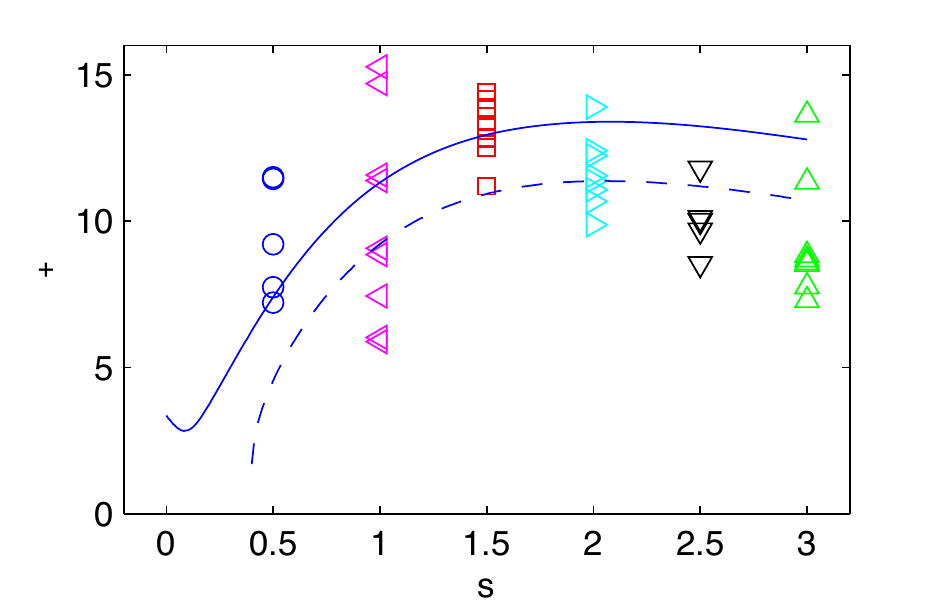}
\caption{%(TOP) Plot of $\Gamma$, \note[LK]{I don't think that $\Gamma$ is defined.} calculated from
%  Eq.~\eqref{eqn:GammaSqr} and preceding equations, with (red) and
%  without (blue) the tip contribution included. As discussed in the
%  text, ~\note[LK]{I don't really find this in the text; I added $\Gamma$ there, but it seems
%  we should also define $\lambda_+$ and possibly include the equations mentioned below, or remove
%  a reference to them?}
%  $\Gamma>0$ corresponds to a rotational instability, which
%  occurs at $s\approx 0.2$ with the tip included and $s\approx 0.4$
%  without the tip included. Thus, $s=0$ intruders should be stable,
%  and all other intruders should be unstable, which is consistent with
 % data presented here. (BOTTOM) 
 A plot of all measured values of
  $\lambda_+$ versus the aspect ratio, $s$. 
  Each data point represents
  a trajectory with sufficient angular deviation (i.e., it has at
  least 3000 data points where $\theta>10^\circ$), where wee measure
  the growth rate as shown by the linear fits imposed on each
  trajectory in Figure~\ref{fig:rotations}. Also plotted is the
  prediction for $\lambda_+$ from the collisional model with (solid line) and without (dashed line) the
  contribution from the intruder tip (see Ref.~\cite{Clark2014} for details).}
\label{fig:GammaInstability}
\end{figure}

The model also provides information on the torque which is exerted on
the intruder. If a non-circular intruder is dropped with a perfect
vertical orientation, then the net torque is zero (discounting the
possibility that instantaneously, the torque may not be zero, due to
fluctuations associated with collisions.)  However, if the inclination
of the vertical axis of an intruder is not vertical, then there is a
net torque, which may be stabilizing, i.e. turning the intruder back
towards vertical, or destabilizing, i.e. tilting the intruder further
from vertical. The details of the intruder boundary shape and the
position of its center of mass both play a role in determining
stability. The net torque on the intruder yields a
second-order-in-time differential equation for the angle, $\theta$
which measures the inclination from vertical. With some approximations
\cite{Clark2014}, this equation can be converted to a second order
differential equation for $\theta(z)$. Stability of the intruder to
rotation is then determined by the eigenvalues of the characteristic
roots of this equation.  Any positive eigenvalue implies instability,
and in such a case, the most positive eigenvalue, $\lambda_+$,
dominates the evolution of $\theta$: $\theta (z) \propto e^{\lambda_+
  z}$. In Fig.~\ref{fig:rotations}, we show data for unstable cases,
which indicate an exponential growth of $\theta$. From our collision
model, we calculate the corresponding growth rate, where now there are
no longer any adjustable parameters. As shown in
Fig.~\ref{fig:GammaInstability}, the measured and predicted growth
rates are in reasonable agreement. For details of the calculations,
see~\cite{Clark2014}.

%
%\begin{figure} \centering
%\includegraphics[trim=0mm 2mm 0mm 0mm, width=0.9\columnwidth]{finaldepth.eps}
%\includegraphics[trim=0mm 0mm 0mm 6mm, width=0.9\columnwidth]{stoptimes.eps}
%\caption{(color online.) Plots of final depths (top) and stopping times (bottom) for all intruders as a function if the kinetic energy at impact.}
%\label{fig:stop-times-and-depths}
%\end{figure}

\section{Impacts at higher speed}

Until this point, we have focused on the regime where the intruder speed is much smaller
than the speed of force transmission in the granular material. In fact, we have taken this
fact for granted in the collisional model presented in the previous section, since we assume the collisions are randomized and point-like over the 
leading edge of the intruder. However, we have not considered what sets the velocity scale for
these grain-grain collisional processes which lead to the primary drag on the
intruder for the previous experiments. One might expect that if the intruder speed became similar to the
grain-grain force transmission speed, the results presented in the previous sections may no longer be valid. As the intruder's and granular force transmission 
speeds become comparable, how does the nature of the force response change, both at the grain scale and macroscopically? For more
information on the material presented in this section, see~\cite{clark2015nonlinear}

We distinguish three relevant speeds during impact: the speed of the
intruder, $v$, the speed of sound in the material from which the
grains are made, $v_b$, and the speed of the propagating signal along
the strong force network, $v_f$. The speed of information propagation
along the force network is set by the force law for interactions
between grains. In general, this law does not depend linearly on the
compression of the interacting grains. Rather, if the distance between
the centers of two grains, $x = R_i + R_j -\delta$, is smaller by
$\delta$ than the sum of their radii, the force typically has the form
\begin{equation}
f = k \delta^{\alpha}
\end{equation} 
where $\alpha = 3/2$ is the Hertz exponent if the particles are
elastic spheres. By direct measurement, we typically find for our
photoelastic disks $\alpha \simeq 1.4$. The constant $k$ depends on
the elastic properties from which the materials are made. These same
elastic properties determine the bulk material sound speed, $v_b$. We
note, that although purely two-dimensional calculation with linear
elasticity predicts $\alpha = 1$ for disks, we always measure $\alpha
\simeq 1.4$ for typical disks used in these experiments.  This finding
may be due to the fact that the disks can expand in the third
dimension, to imperfections in disk surface, or possibly to other
effects.

As force is carried down a force chain, the time scale for transmission
from one grain to the next is the collision time, $t_c$, which is set
by the force law:

\begin{equation} 
t_c =d v_0^{\frac{1-\alpha}{1+\alpha}} v_b ^\frac{-2}{\alpha +1} C(\alpha),
\label{eqn:tc}
\end{equation}
where $C(\alpha)$ depends on $\alpha$ but otherwise is a
constant. Note that $t_c$ is much longer than the time $d/v_b$ for
stresses to propagate a distance $d$ at $v_b$, where $v_b=(E^*/
\rho)^{1/2}$ is the bulk sound speed, and $d$ is the mean grain diameter.
Here, $E^*$ is an effective elastic modulus, and $\rho$ is the mass density of the grains (which we
determine experimentally); see Experimental Techniques section below for more information.

The speed of force propagation down a chain is set by $t_c$ and the
diameter, $d$ of a grain:
\begin{equation}
v_f \propto d/t_c
\end{equation}
or,
\begin{equation}
v_f \propto \frac{d}{t_c} = v_0^\frac{\alpha -1}{\alpha +1} v_b ^\frac{2}{\alpha +1} \left[C(\alpha)\right]^{-1},
\label{eqn:v-scaling}
\end{equation}
If energy is conserved, then $t_c$ is given by
eq.~\ref{eqn:tc}. However, a similar scaling still applies even if
energy is not conserved, as long as the energy loss is not too great.

This leads to a scaling relation involving ratios of $v_f$, $v_0$, and $v_b$:
\begin{equation}
\frac{v_f}{v_b} \propto \left(\frac{v_0}{v_b}\right)^\frac{\alpha -1}{\alpha +1} .
\label{eqn:scaling}
\end{equation}
which forms the basis for understanding the speed of front propagation
as a function of impact speed and material properties. Note that Gomez
et al.\cite{Vitelli2012} have obtained a similar relation for systems
of frictionless grains, where they have invoked equipartition between
potential and kinetic energy of grains. The present argument, is based
on the fact that in systems of frictional grains, for which there is
no well defined conserved energy, the force is carried along
sometimes-tenuous strong force networks, which are strongly loaded and
then unloaded by impact.

\begin{figure} 
\raggedright (a)\\ \includegraphics[trim=20mm 0mm 20mm 0mm, clip, width=0.8\columnwidth]{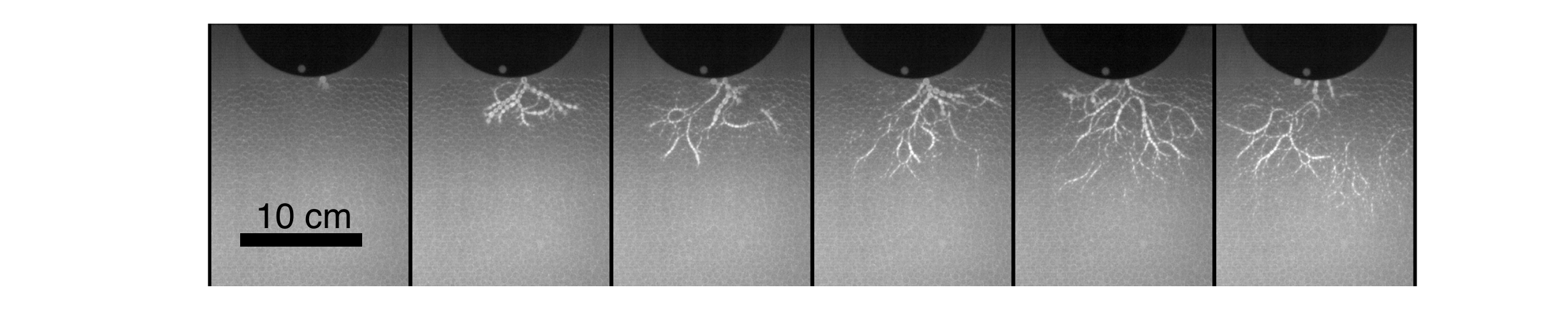} \\
\raggedright (b)\\ \includegraphics[trim=20mm 0mm 20mm 0mm, clip, width=0.8\columnwidth]{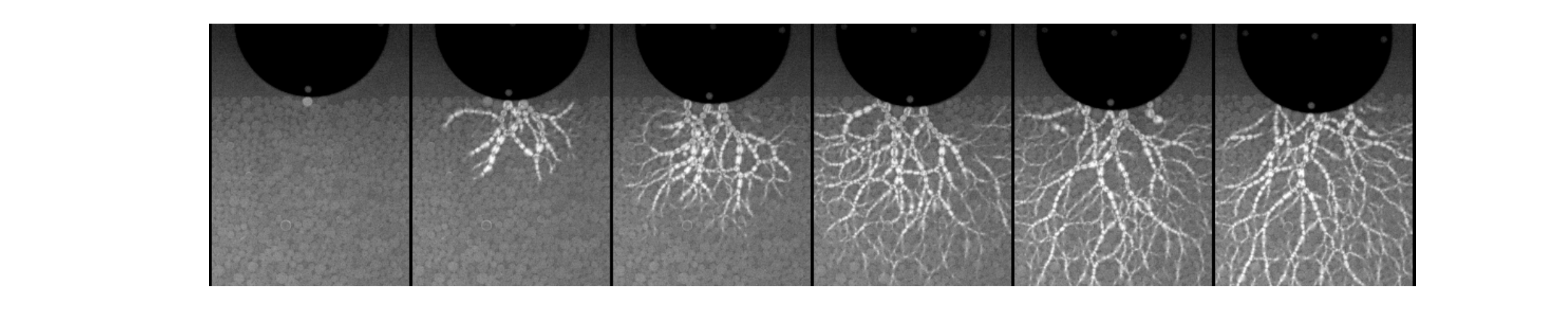} \\
\raggedright (c)\\ \includegraphics[trim=20mm 0mm 20mm 0mm, clip, width=0.8\columnwidth]{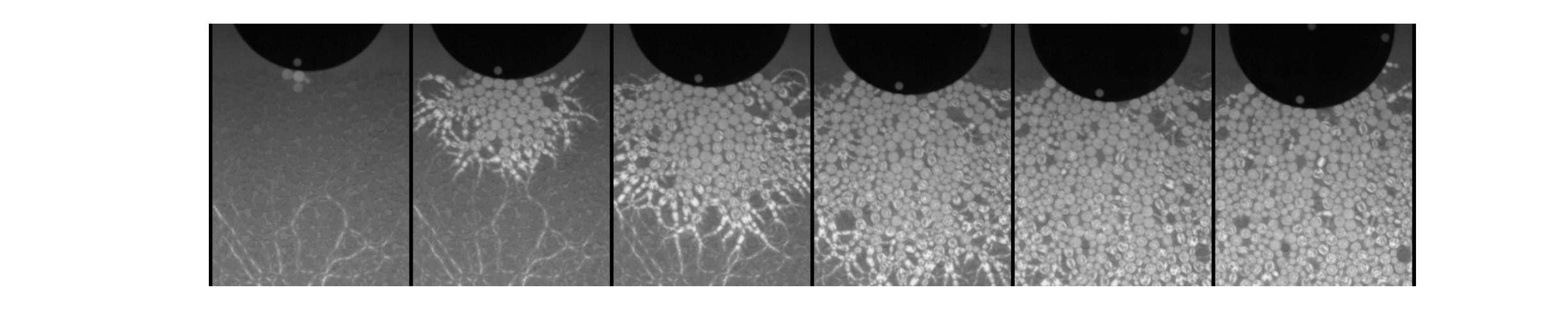} \\
\caption{Frames showing force propagation after three impacts with $v_0\approx 5$~m/s. (a) The hardest particles ($M'\approx 0.1$) correspond to fast, chain-like force propagation. (b) Forces for intermediate stiffness particles ($M'\approx 0.3$) are more dense spatially, but still relatively chain-like. (c) The softest particles ($M'\approx 0.6$) show a dense force structure which propagates with a well defined front.}
\label{fig:vary-M}
\end{figure}

To significantly vary the ratio of relative speed of impact, $v_0$, to
the speed of propagation along the strong network, $v_f$, i.e. $M' =
v_0 t_c/d \sim v_0/v_f$, we use materials of three different
stiffnesses. By these means, we can vary $M'$ by over two orders of
magnitude. In Fig.~\ref{fig:vary-M}, we contrast frames from impacts
at low, moderate and high $M'$. In evaluating $M'$, we use the
assumption, justified below, that $v_f \propto d/t_c$. A key
observation is that for the lowest $M'$, the force is carried
primarily over a filimentary force network (bright particles in the
upper image of Fig.~\ref{fig:vary-M}.) As $M'$ increases, the strong
force network becomes increasingly dense, so that for the largest $M'$
case, the propagating force network is spatially dense and not
filimentary.

\begin{figure} 
\centering
\raggedright (a) \\ \centering \includegraphics[trim = 0mm 0mm 0mm 2.5mm, clip, width=0.7\columnwidth]{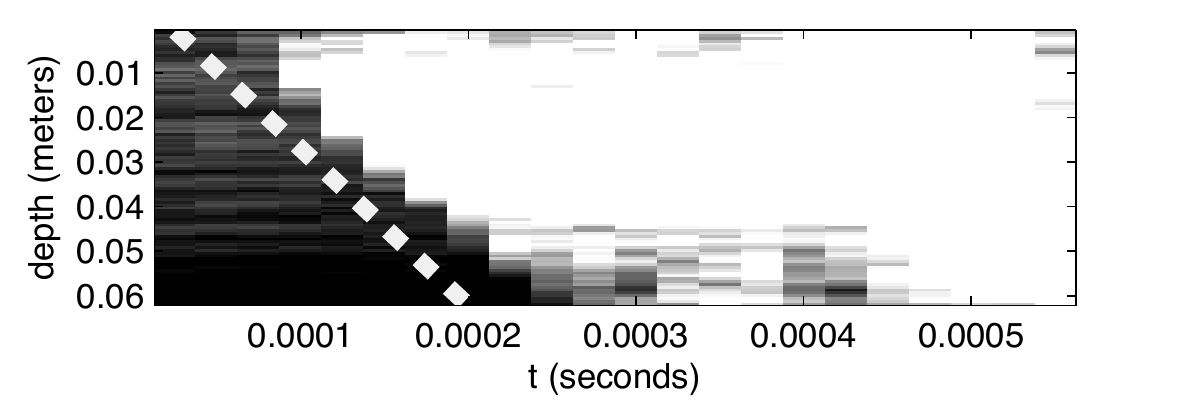} \\
\raggedright (b) \\  \centering \includegraphics[trim = 0mm 0mm 0mm 2.5mm, clip, width=0.7\columnwidth]{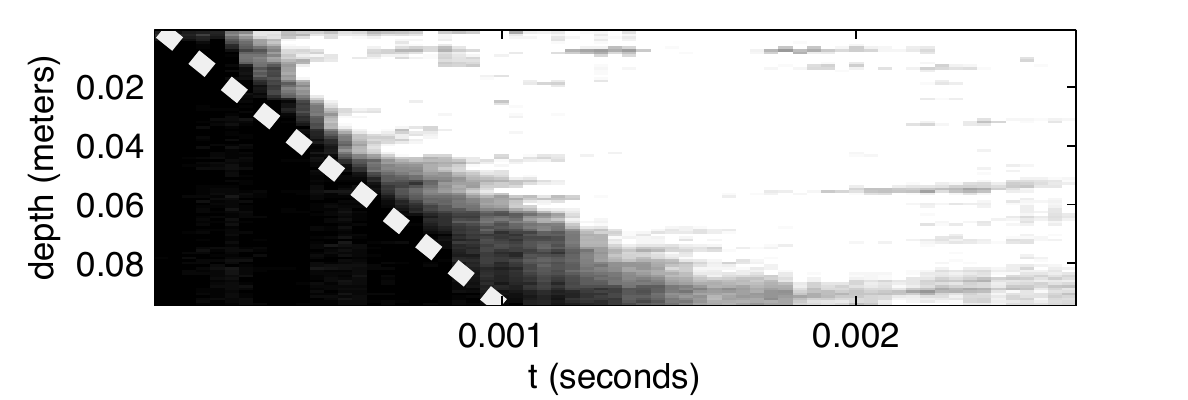} \\
\raggedright (c) \\  \centering \includegraphics[trim = 0mm 0mm 0mm 2.5mm, clip, width=0.7\columnwidth]{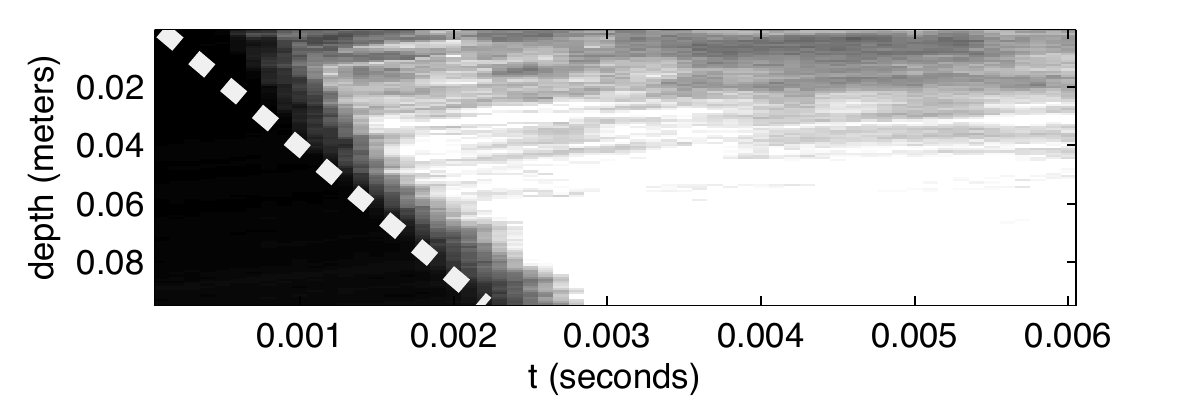} \\
\raggedright \hspace{2in} (d) \hspace*{1.5in} (e) \\ \centering
\includegraphics[trim = 0mm 0mm 3mm 2.5mm, clip, width=0.25\columnwidth]{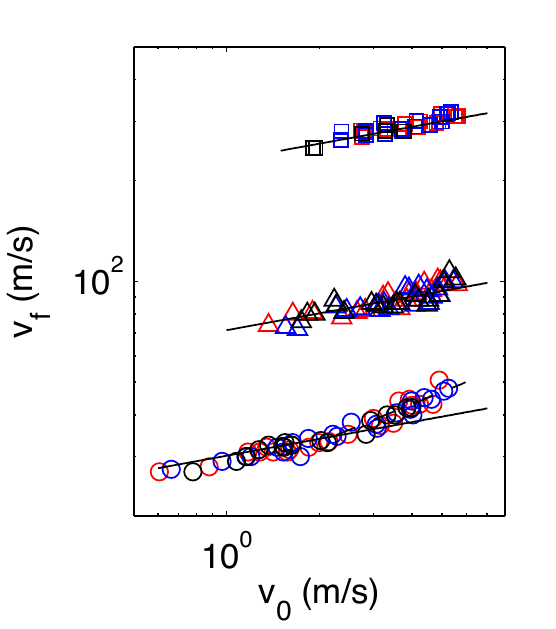}
\includegraphics[trim = 0mm 0mm 4mm 2.5mm, clip, width=0.25\columnwidth]{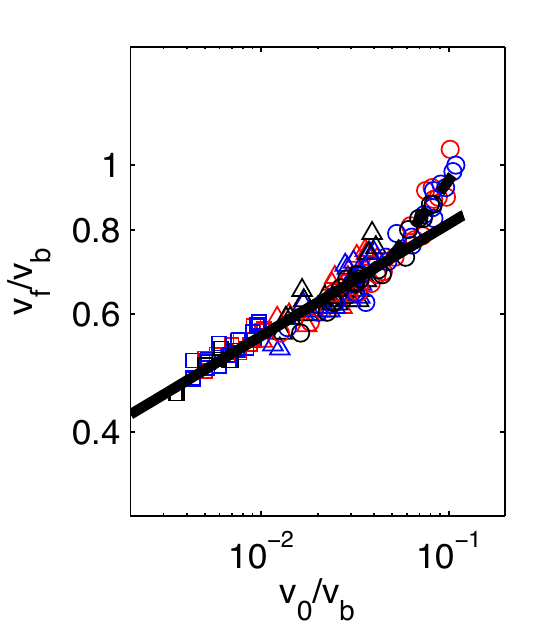}

\caption{(a)-(c) Space-time plots show forces propagating for impacts shown in Fig.~\ref{fig:frames} (see text for details). Dashed lines indicate $v_f$. (d) These are plotted versus $v_0$, where symbol shapes represent different particle stiffness (squares/triangles/circles represent hard/medium/soft particles), and symbol colors represent different intruder diameters (red/blue/black for 6.35/12.7/20.32~cm). (e) When velocities are normalized by the bulk sound speed these data collapse onto a single curve. The solid fit line corresponds to Eq.~\eqref{eqn:scaling}. The dashed fit line corresponds to collective stiffening of the soft particles for large deformations, as discussed in the text.}
\label{fig:spacetime-plots}
\end{figure}

We coarse-grain frames from impact events to obtain a coarse-grained
space-time plot of the photoelastic response as a function of time and
depth under impacting intruders. Fig.~\ref{fig:spacetime-plots} shows
the space-time plots corrsponding to the impacts shown in
Fig.~\ref{fig:vary-M}. This smoothing averages over force networks,
although these are an essential feature of the signal propagation. We
then plot the measured front speeds vs. the initial impact speed,
$v_0$, in the lower left part of Fig.~\ref{fig:spacetime-plots}. The
data fall on separate curves for each type of particle material, and a
fit to these data is in good agreement with
Eq.~\ref{eqn:v-scaling}. All of these data can be collapsed onto a
single curve, as suggested by Eq.~\ref{eqn:scaling}.

\begin{figure} \centering
\hspace{0.5in}\raggedright (a) \hspace{3in} (b) \\ \centering
\includegraphics[trim = 0mm 0mm 0mm 2.5mm, clip, width=3in]{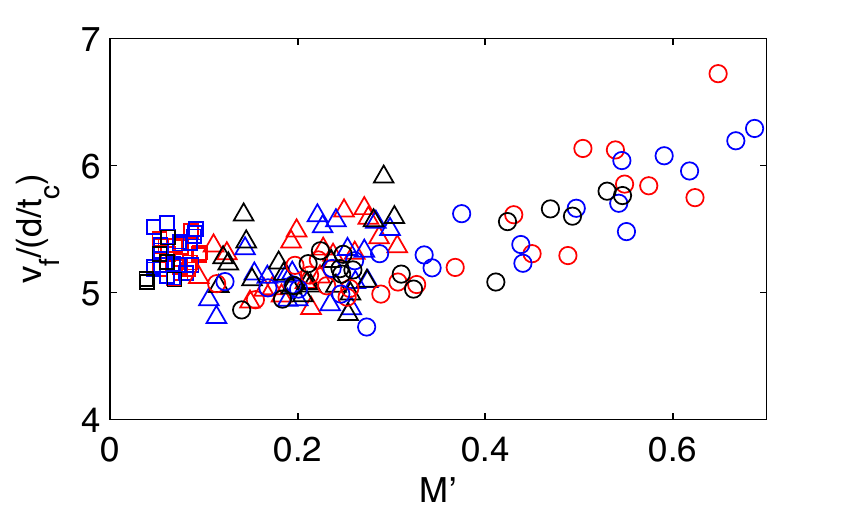}
\includegraphics[trim = 0mm 0mm 0mm 3mm, clip, width=3in]{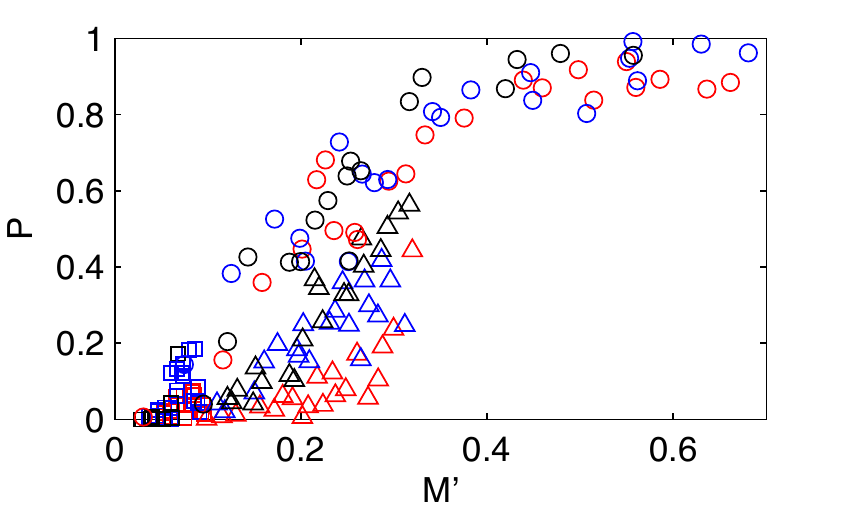}
\caption{(a) Ratio of measured propagation speed, $v_f$, to $d/t_c$, as a function of $M'$. The propagation speed for $M'<0.4$ is $v_f\approx 5.3{d}/{t_c}$. For $M'>0.4$, the force propagation speed is faster than this relation. (b) Participation ratio, $P$, plotted as described in the text, as a function of $M'$. The spatial density of the forces saturates ($P\approx 1$) when $v_f$ departs from $v_f \propto d/t_c$ for $\alpha = 1.4$.}
\label{fig:M-prime-scaling}
\end{figure}

There is a departure from the power-law scaling of
Eq.~\ref{eqn:v-scaling} and Eq.~\ref{eqn:scaling} at the highest $M'$,
which also corresponds mostly to impacts into the softest material. We
show an alternative characterization of this effect in
Fig.~\ref{fig:M-prime-scaling}, top, in the form of a plot of
$v_f/(d/t_c)$ vs. $M'$. Note that at high $M'$ the forces are no
longer carried on a filamentary network, and for the strong forces
involved in this case, extra contacts can form, strengthening the
material. We further characterize this effect also through a
participation ratio, $P$, defined as the fraction of particles
carrying large forces. $P$ rises steadily with $M'$, and begins to
approach an asymptotic value of order unity at roughly a similar value
of $M'$ where the ratio $v_f/(d/t_c)$ begins to rise. We note
  two phenomena connected to the departure at high $M'$ from the power-law scaling in Eqs.~\ref{eqn:v-scaling} and~\ref{eqn:scaling}. One is that the number of
  particles in the strong network grows with $M'$, since the force
  transmission slows relative to the rate at which new parts of the
  network are created. The other phenomonon is that under strong
  enough forcing, new contacts form, which strengthens and homogenizes
  the force structure.

\section{Simulation results for penetration depth at large M'}

Numerical simulations provide a useful tool in the study of granular
materials, in particular since, in a simulation, vs. an experiment, it
is much easier than in an experiment to change the system parameters,
such as friction, particle size, particle stiffness. The discussion
given here is focused on the influence of system parameters on the
penetration trajectory. More in-depth analysis can be found
in~\cite{pre12_impact}, and the details about the simulations are
given in the Techniques section. We note that simulations using a
nonlinear force model ($\alpha \ne 1$) are in progress and will allow
more detailed comparision to experiments.

\begin{figure} \centering
 \centering
 \includegraphics[width=2.5in]{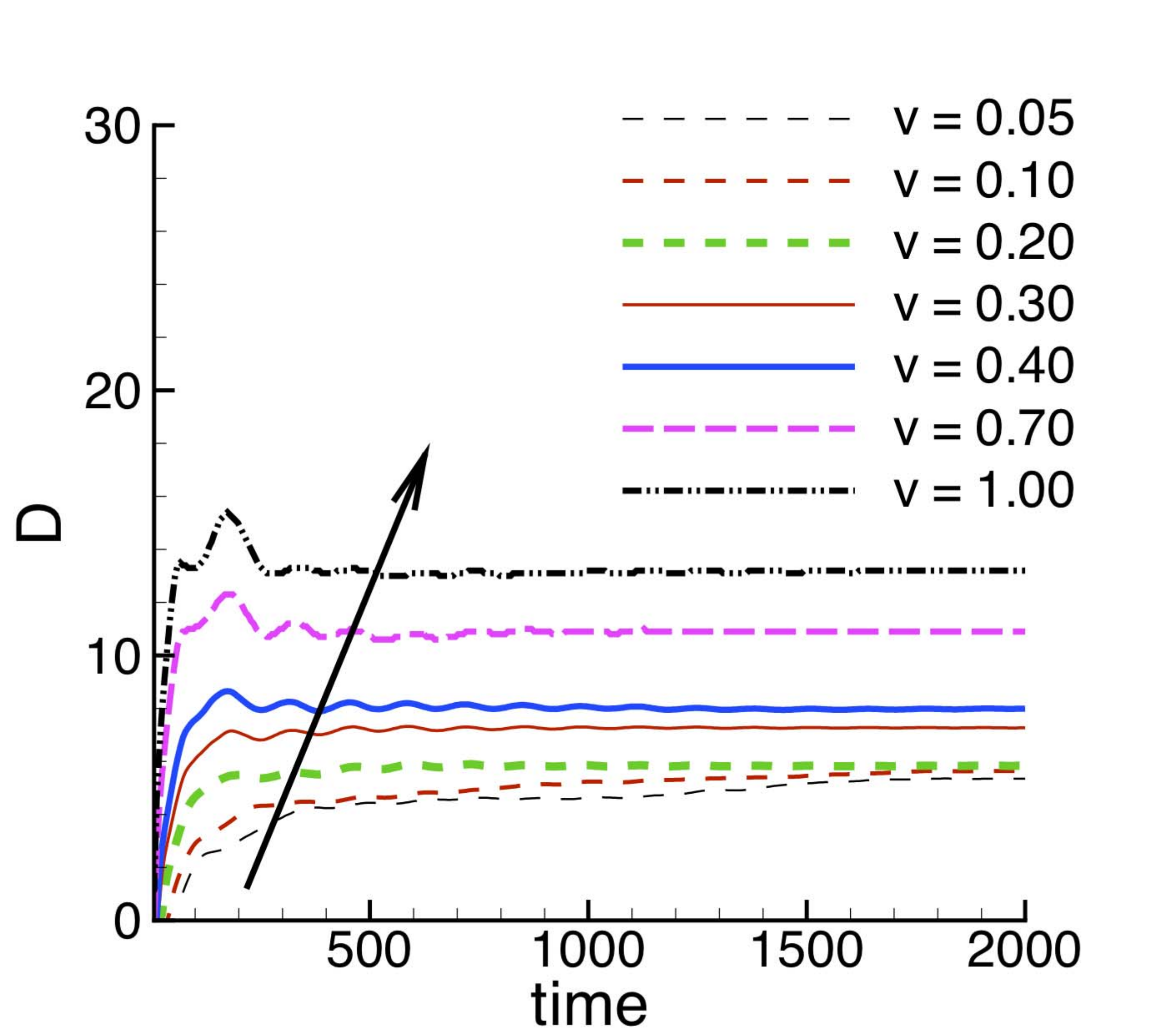}
\caption{Penetration depth of $D_i=10$ intruder impacting with
  different speed. Here we use $r = 0.2$, $k_n = 4\cdot 10^3$, $k_t =
  0.0$, $e=0.5$, $\mu = 0.5$. Material properties of the intruder
  are the same as of the granular particles. The arrow shows the
  direction of increasing impact speed.}
\label{fig:basic}
\end{figure}

We start by considering a randomly packed system with particles
characterized by polydispersity $r=0.2$, so that the particle sizes
are uniformly distributed in the range $d(1\pm r)$, where $d$ is the
average particle diameter, $\mu = 0.5$ is the static friction
coefficient, $e = 0.5$ is the coefficient of restitution. The
simulations discussed here only have kinetic friction (viscous
damping) only, so that tangential springs at the point of contact are
removed ($k_t=0.0$), as disucssed in the Techniques section.  The
normal forces are modeled by linear springs, so that $\alpha = 1$
here.  The intruder is a disk, with diameter $D_i= 10$ in units of
$d$, and otherwise possessing the same material properties as the
granular particles. Figure~\ref{fig:basic} shows the time evolution of
$z(t)$ of an intruder impacting the granular bed with one of seven
different speeds, ranging from $0.05$ to $1$, scaled by $d/\tau_c$,
where $\tau_c$ is a typical binary collision time between the
particles (impact speed independent for the linear force model
considered here).

The key results presented in Fig.~\ref{fig:basic} are as follows. As
expected, slower initial impact speeds, $v_0$, create shallow craters;
specifically, the penetration depth is less than the intruder's own
diameter. By contrast, intruders of higher speeds are entirely
submerged in the granular bed. For the larger impact velocities, we
find an overshoot in the penetration depth, i.e., the intruder
rebounds towards the surface of the granular layer, as also observed
experimentally~\cite{Goldman2008, Daniels2004}.  Next, we proceed to
analyze the influence of the properties of the granular system on the
penetration depth.

\begin{figure} \centering
 \centering
\includegraphics[width=2.5in]{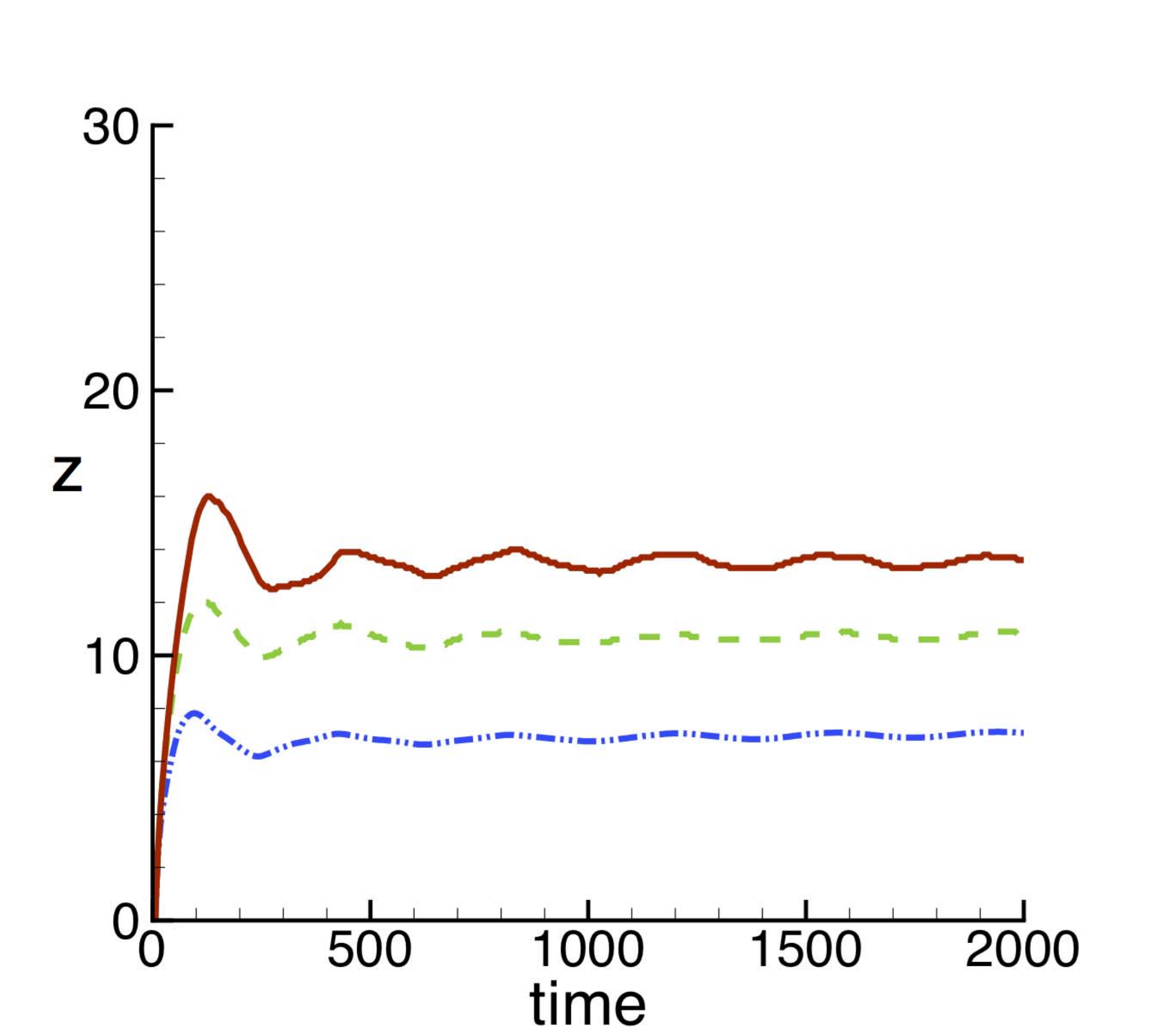}
\caption{Penetration depth of an intruder in systems with varying
  coefficient of restitution, $e = 0.9,~0.5,~0.1$ shown by red
  (solid), dashed (green) and blue (dash-dot) lines, respectively.
  Here, $v=0.7$, the system size is $W=200$, $L=200$; the other
  parameters are as in Fig.~\ref{fig:basic}. }
\label{fig:200x200_en0.9}
\end{figure}

 Figure~\ref{fig:200x200_en0.9} shows the influence of elasticity of
 the granular particles, measured by the coefficient of restitution,
 $e$. A large coefficient of restitution leads to a significantly
 deeper penetration, as might be expected, since the energy loss is
 reduced relative to a lower restitution coefficient. Interestingly,
 while a decrease of $e$ reduces the depth of penetration, it does not
 remove the overshoot of the $z(t)$ curve.  We will see below that a
 different behavior results when the frictional properties of granular
 particles are modified.
 
The influence of friction between the granular particles on the
trajectory, $z(t)$, in particular, and on the response of the granular
material to an intruder in general is not immediately obvious. For
example, in considering the response of a system to a point force, it
has been found that friction plays a role in determining how forces
and stresses propagate through the system~\cite{Goldhirsch2005}. Of
course, a response to an intruder is expected to be more complicated
since it leads to a large scale rearrangement of granular particles,
which is not expected in a response to a localized (small) point
force. Indeed, it has been suggested that friction is not necessarily
crucial in understanding this response~\cite{Seguin2009}.
 
\begin{figure} \centering
 \centering
\includegraphics[width=2.5in]{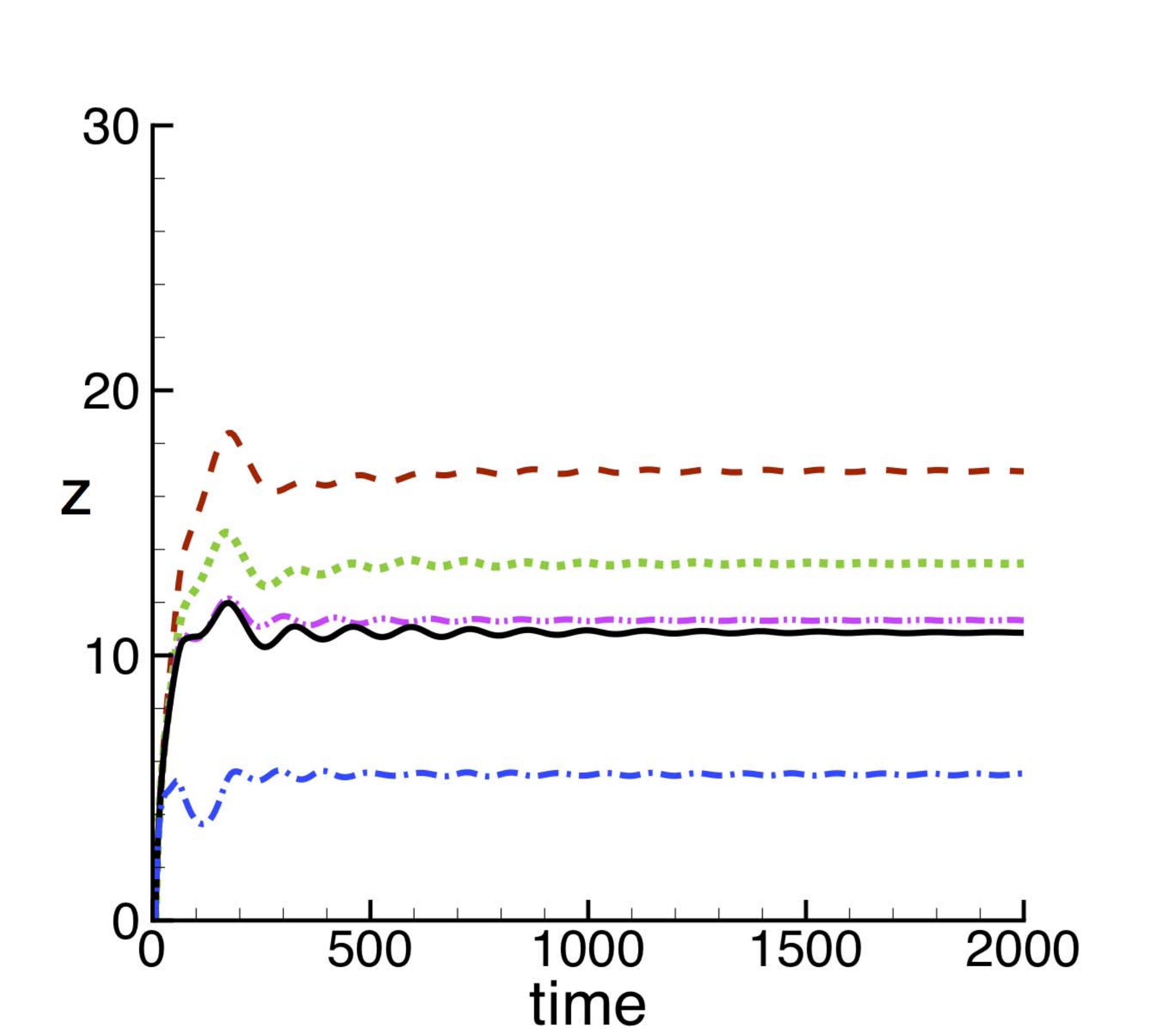}
\caption{Penetration depth for different friction models and Coulomb
  thresholds; here we show the results as follows (top to bottom):
  $\mu = 0$ (red dashed); $\mu = 0.1,~k_t = 0.0$ (green dotted);
  $\mu=0.1,~k_t = 0.8$ (pink dash-dot-dot), $\mu=0.5,~k_t = 0.0$
  (black solid); $\mu=0.5,~k_t = 0.8$ (blue dash-dot). Here $v=0.7$,
  the other parameters are as in Fig.~\ref{fig:basic}.}
\label{fig:friction}
\end{figure}

 To illustrate the influence of friction, we consider two effects:
 first, the effect of the friction model, and second, of Coulomb
 threshold. Figure~\ref{fig:friction} shows the corresponding
 results. We find that having a model with static friction leads to a
 significantly smaller penetration depth (blue dash-dotted curve in
 Fig.~\ref{fig:friction}) than a model without static friction,
 particularly for a large Coulomb threshold. For a smaller Coulomb
 threshold, the influence of static friction is weaker, and the
 response of the system in that case turns out to be similar to the
 one obtained using kinetic friction only (compare green dotted and
 pink dash-dot-dot curves in Fig.~\ref{fig:friction}). Furthermore, an
 `overshoot' in the intruder depth may be removed in the case of
 (strong) static friction. This is one significant difference between
 the influence of friction and inelasticity on the intruder's
 dynamics: in the case of increased inelasticity (smaller coefficient
 of restitution), we still find an `overshoot', as in
 Fig.~\ref{fig:200x200_en0.9}.

To summarize this section, the simulations show that the material
properties of granular particles (in particular elasticity and
friction) play a significant role in determining dynamics of the
intruder in the regime considered. Future work shall discuss the
influence of these and other parameters in the regimes relevant to
experiments discussed in this review.

\section{Techniques for Experiments and Simulations}

\subsection{Experimental Techniques}

\subsubsection{Experimental Apparatus} 
The experimental apparatus\cite{Clark2012,Clark2013,Clark2014}
consists of two Plexiglas sheets (0.91~m $\times$ 1.22~m $\times$
1.25~cm) separated by a thin gap (3.3~mm), and filled by photoelastic
disks (thickness of approximately 3~mm). We construct the intruders
from bronze sheet (bulk density of 8.91~g/cm$^3$ and thickness of
0.23~cm).  We drop the intruders from a height $H \leq
2.2$~m, through a shaft connected to the top of the thin gap
containing the particles. This produces impact speeds $v_0 \simeq
(2gH)^{1/2}$. A Photron FASTCAM SA5 records results at frame rates of
10,000, 25,000, and 40,000 frames per second for the soft,
intermediate, and hard particles, respectively. We measure the
intruder velocity at impact by locating the intruder with a
circular Hough transform at each frame and computing the velocity with a
numerical derivative. We use the intensity of the photoelastic image
at each frame to compute space-time plots to study the spatial
structure of forces and measure $v_f$. We estimate
an approximate uncertainty in the propagation velocity for each impact
to be $\pm5\%$.

We use particles made from three photoelastic materials, each with a
different stiffness. The softest material is polyurethane sheet
(Precision Urethane) with a hardness rating of Shore 60A, cut into
disks of 6 mm and 9 mm diameter. The second material is a stiffer
polyurethane sheet (Precision Urethane) with hardness rating Shore
80A, cut into disks with diameters of 6 mm and 9 mm. The stiffest
material is PSM-1, (Vishay Precision Group) cut into disks with
diameters of 4.3~mm and 6~mm. We measure the force versus compression
for single particles in separate experiments. We find that, for
compression between two plates (similar to particle compression in
force chains), a single scaling relation captures the behavior of all
types of particles
\begin{equation}
f = E^* w d \left(\frac{\delta}{d}\right)^\alpha,
\label{eqn:force-law}
\end{equation}
where $f$ is the compression force, $w$ is the particle thickness, $d$
is the particle diameter, $\delta$ is the displacement, and $\alpha
\approx 1.4$ for all particles. The magnitude of the force law is the
effective Young's modulus, $E^*$, which is set by properties of the
bulk material, including effects from the Poisson ratio. We measure $E^*
\approx 3$~MPa for the Shore 60A material, $E^* \approx 23$~MPa for
the Shore 80A material, and $E^* \approx 360$~MPa for the PSM-1
material.

\subsubsection{Photoelastic Techniques}
The experiments described here use photoelastic particles. as used by a number of researchers\cite{Howell1999,Atman2005,Geng2001,Majmudar2005,Majmudar2007,Krim09,Geng2001PRL,Ren2013,Bi2011,Puckett2013, Zhang2010,Geng2003,Rajchenbach2000,Tordesillas2009, Dantu1968,Wakabayashi1950,Drescher1972,Drescher1976,Utter2008,Bassett2012}. The important feature associated with particles made from a
photoelastic material is that they yield particle-scale
force information. 

If a disk or other quasi-2D object is
illuminated by an incident beam of intensity $I_0$ and placed between
crossed circular polarizers, then the fraction of light intensity that
emerges for a ray that has traversed this disk is given by an
expression that connects the local shear stress, $\tau = \sigma_2 -
\sigma_1$:
\begin{equation}
I/I_0 = \sin[(\sigma_2 - \sigma_1) C T/\lambda].
\label{eq:photo-el}
\end{equation} 
The $\sigma_i$ are the principal stresses in the object, evaluated at
the location of the light ray; $C$ is the material-dependent stress
optic coefficient; $T$ is the object thickness along the direction of
transmitted light; $\lambda$ is the wavelength of the light.

We use an empirical approach, similar to that described by Howell et
al.\cite{Howell1999}. The idea is that contact forces acting on a
grain create stress fields inside the grain which cause the phase
variable, $(\sigma_2 - \sigma_1) C T/\lambda$, inside the sine
function of Eq.~\ref{eq:photo-el} to wrap through multiples of $\pi$.
Where the phase variable is an integer multiple of $\pi$, the
corresponding transmitted image region is a dark, and where the phase
is an odd multiple of $\pi/2$ it is bright. In a photoelastic image of
such a grain, increasing applied contact forces increases the stresses
(both pressure and shear stress) within the grain, and leads to an
increasing density of light/dark fringes. Where these fringes appear
in the image depends on the wavelength/color of the light. Since the
pressure is a reflection of the mean normal forces on a particle,
hence the internal stress, it is possible to calibrate a measure of
the fringe density (or sometimes total intensity) against the pressure, $P$. 

In these experiments, the forces due to impact cause the particles
experiencing large forces to appear as bright in a photoelastic image,
but in general we do not resolve fringes. For particles made from the
harder materials, the brightness of a particle in an image suffices to
yield the force acting on it. For the softest material, multiple
fringes can appear in a particle. In this case, particles experiencing
the largest forces may appear slightly dimmer than particles
experiencing somewhat less force, since for the largest forces, the
apparent particle brightness is averaged over light and dark fringes.

\subsubsection{Intruder Shape}

The circular intruders are machined from bronze sheet (bulk density of
8.91~g/cm$^3$ and thickness of 0.23~cm) into disks of diameters $D$ of
6.35~cm, 10.16~cm, 12.7~cm, and 20.32~cm.

We used four ogive intruders. The ogives consisted of a
continuous piece of material, where the leading part was a half-ellipse
truncated along the minor axis, with semi-major axis $a$ and
semi-minor axis $b=D/2$, terminated by a rectangular tail of length
$L$. We used three different ellipses, with $a/b=1$ (half-circle),
$a/b=2$, and $a/b=3$. We kept the width of the ogives  constant,
$D=9.3$~cm, and we varied $L$  to keep the intruder mass constant
($L=b=4.15$~cm for $a/b=3$ case, longer for other ogives). By keeping
the width and mass constant, we isolate shape effects. Additionally,
we used one smaller ogive with $a/b=1$, $b=3$~cm, and $L=7.7$~cm.

The triangular-nosed intruders are comprised of a downward-pointing
isosceles triangle, symmetric about the vertical axis, with opening
angle $2\alpha$, attached to a rectangular tail of the same width as
the base of the triangle, $W=9.65$~cm. The noses of these intruders
are clearly evident in Fig.~\ref{fig:frames}. The length of the tail
is varied to keep the total area, $A=0.0107$~m$^2$, and hence, mass,
$m=0.219$~kg, constant for different opening angles of the nose (for
reference, the $s=3$ intruder has a tail which is 3.81~cm long). Thus,
the intruder nose has a constant magnitude slope $s=\tan^{-1}\alpha$,
except at the tip, which is rounded with a radius of about
1.5~mm. Note that this is smaller than the particle radii, which are
greater than and 4.3~mm. The $s=3$ intruder is turned upside-down and used as
the $s=0$ intruder.

\subsection{Data Processing}
At each frame, we use distinguishing features of the intruder to
locate its center of mass relative to the initial point of impact and
its angular position relative to the vertical direction with errors of
less than 1 pixel (0.5~mm). This yields the intruder trajectory, and
the rotation angle, $\theta$. By discrete differentiation, combined
with a low-pass filter, we obtain the velocity and acceleration. Using
the data for $z(t)$, $v(t)$, and $a(t)$ for many different
trajectories with varying initial velocities, we fit to a force-law
model such as Eq.~\eqref{eqn:forcelaw}. This allows us to
experimentally measure $f(z)$ and $h(z)$ for each intruder. This
process is described in detail in \cite{Clark2013}.

\subsection{Simulation Techniques}
\label{app:simulations}

We consider a rectangular domain in two dimensions with gravity. The
particles are polydisperse discs, with their diameters varying
randomly in a range $\pm r$ about the mean. 
The particle-particle and particle-wall interactions are modeled using the soft-sphere approach
that includes friction and rotational degrees of freedom. We solve
the following equations of motion for each particle:
\begin{eqnarray}
m_i\frac{d^2\mathbf{r}_i}{dt^2} &=& \mathbf{F}^n_{i,j}+    \mathbf{F}^t_{i,j}  +  m_i\mathbf{g},\nonumber
\\
I_i\frac{d\boldsymbol{\omega}_i}{dt} &=&
-\frac{1}{2}d_i\mathbf{n}\times\mathbf{F}^t_{i,j}.
\label{eq:motion}
\end{eqnarray}
The normal force is given by 
\[{\bf F}^n_{i,j} =  \left[ k_n x - \gamma_n \bar m {\bf v}_{i,j}\right ] {\bf n},\]
where $r_{i,j} = |{\bf r}_{i,j}|$, ${\bf r}_{i,j} = {\bf r}_i - {\bf
  r}_j$, and the normal direction is defined by ${\bf n} = {\bf
  r}_{i,j}/r_{i,j}$. The compression is defined by $x =
d_{ave}-r_{i,j}$, where $d_{ave} = {(d_i + d_j)/2}$, $d_{i}$ and
$d_{j}$ are the diameters of the particles $i$ and $j$; ${\bf
  v}_{i,j}^n$ is the relative normal velocity. 

The nondimensional force constant, $k_n$, is related to the dimensional
one, $k$, by $k = k_n mg/d$, where $m$ is the average particle mass,
$d$ is the average particle diameter, and $g$ is Earth's gravity. All
quantities are expressed using $d$ as the length scale, the binary
collision time
\[\tau_c =
\pi \sqrt{d\over 2 g k_n},\] as the time scale, and $m$ as the mass
scale. Then, $\bar m$ is the reduced mass, and $\gamma_n$ is the
damping coefficient related to the coefficient of restitution, $e$, by
$\gamma_n = -2\ln e/\tau_c$~ (e.g.,Kondic~\cite{kondic_99}). We take
$e$ constant and ignore its possible velocity
dependence~\cite{schafer96}. For definitiveness, we typically use the
physical parameters that are appropriate for {\bf the softest}
photoelastic disks~\cite{Geng2003}, in particular $d=4$ mm, $k_n =
4\cdot 10^3$, $e = 0.5$. The parameters entering the force model can
be connected to the physical properties of the material (Young
modulus, Poisson ratio) using the method described in
Kondic\cite{kondic_99}.

%Before proceeding with the discussion of the tangential forces, it is
%appropriate to comment on the presence of two time scales in the
%problem: one is the fast time-scale, $\tau_c$, defined above, which is
%relevant for the processes involving particle collisions, and the
%other slow penetration time scale is $t_s = \sqrt{D_i/g}$, where
%$D_i$ is the intruder's diameter. Here, $t_s$ is proportional to the time for an intruder to
%travel a distance equal to its own diameter, starting from rest in a
%gravitational field. In this work, we will concentrate chiefly on the
%granular dynamics, and therefore $\tau_c$ is the most appropriate time
%scale, and $d/\tau_c$ is the most appropriate (fast) velocity scale.
%The two time scales are related by $\tau_c/t_s = \pi/\sqrt{2 D_i
%  k_n}\ll 1$.

The tangential force is specified by two different models, which can
be conveniently described within the same framework. The basic
approach is based on a Cundall-Strack type of model~\cite{cundall79},
where a tangential spring of zero length is introduced when a new
contact between two particles forms at $t=t_0$. Due to relative
motion of the particles, the spring length, ${\boldsymbol\xi}$, evolves
as
\[ \boldsymbol\xi=\int_{t_0}^t {\bf v}_{i,j}^t~(t')~dt',\]
where ${\bf v}_{i,j}^{t}= {\bf v}_{i,j} - {\bf v}_{i,j}^n$. For long lasting contacts,
$\boldsymbol\xi$ may not remain parallel to the current tangential direction defined by
$\bf {t}={\bf v}_{i,j}^t/|{\bf v}_{i,j}^t|$ (see, e.g,.~\cite{brendel98,laetzel03}); we therefore
define a corrected
$\boldsymbol\xi{^\prime} = \boldsymbol\xi - \bf{n}(\bf{n} \cdot \boldsymbol\xi)$ and
introduce the test force
\[
{\bf F}^{t*} = -k_t\boldsymbol\xi^\prime - \gamma_t {\bf v}_{i,j}^t,\]
where $\gamma_t$ is the coefficient of viscous damping in the
tangential direction (we use $\gamma_t = {\gamma_n/2}$). To keep the
magnitudes of tangential forces smaller than the Coulomb threshold,
specified by $\mu {\bf F}^t$, where $\mu$ is the coefficient of static
friction, we define the tangential force by
\begin{equation}
{\bf F}^t = min(\mu |{\bf F}^n|,|{\bf F}^{t*}|){{\bf F}^{t*}\over|{\bf F}^{t*}|}.
\label{eq:ftan}
\end{equation}
In addition, $\boldsymbol\xi^\prime$ is reduced to the length
corresponding to the value of $|{\bf F}^t|$ as needed. This is a
commonly used model for static friction, with non-zero $k_t$. To be
able to isolate the effect of static friction, we also consider a
commonly used kinetic friction model based on viscous damping, which
is obtained simply by putting $k_t = 0$. Therefore, depending on
whether static friction effects are considered or not, we use either
model 1: $k_t=0.8 k_n$ (the value suggested in~\cite{goldhirsch_nature05}), or
model 2: $k_t=0.0$ (kinetic friction only). The exact value of $k_t$ does not seem to be of relevance
in the present context as long as $k_t \ne 0$. The particles making up the walls
are made very inelastic and frictional, with $\mu=0.9$ and $e=0.1$.

\begin{figure} \centering
\centering
\includegraphics[width=3in]{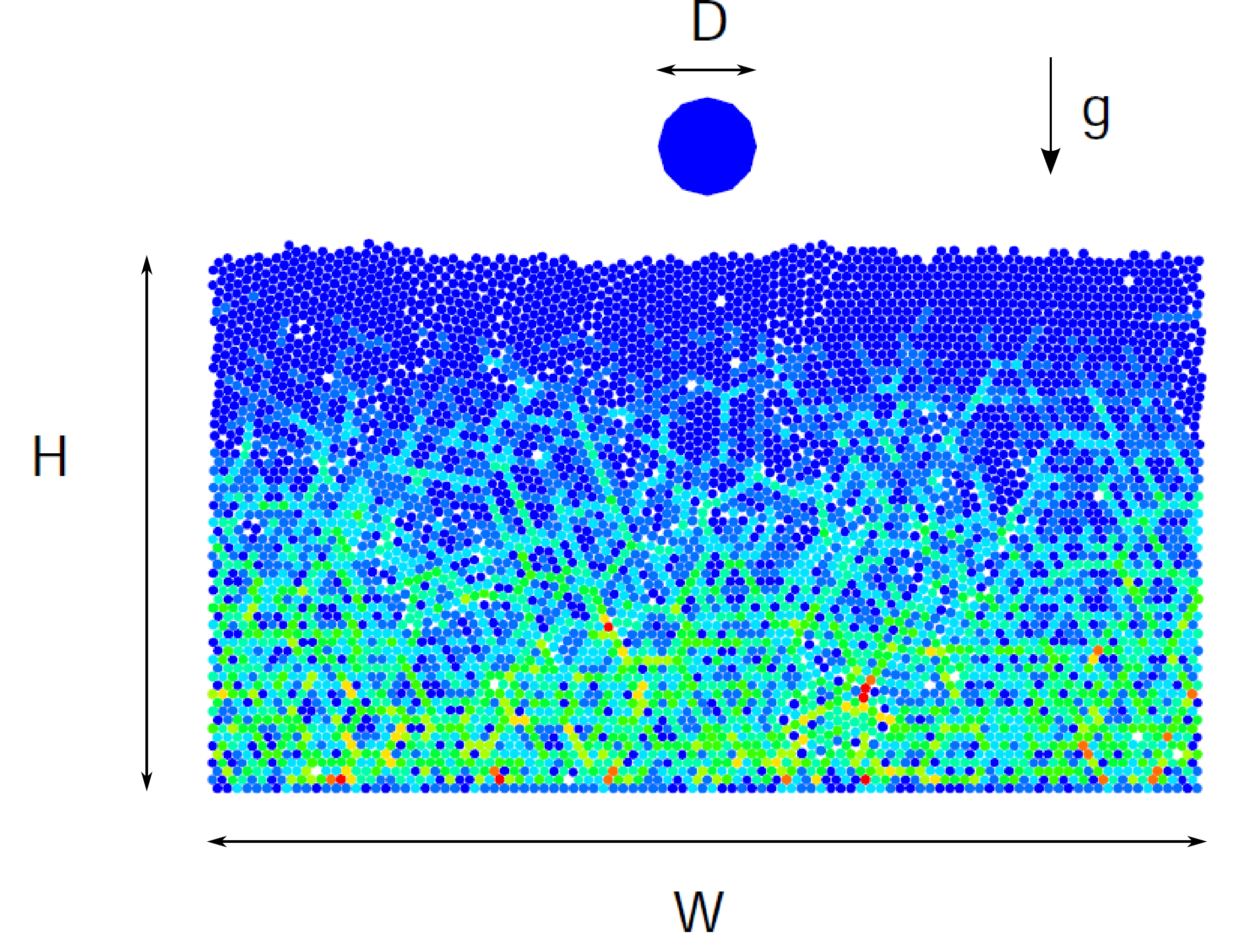}
\caption{System setup. Particles are colored according to the total normal
  force experienced (here only due to gravity).
  }
\label{fig:setup}
\end{figure}

Figure~\ref{fig:setup} shows the system setup. Here, $W$ and $L$ are
the width (depth)  and the length of the granular bed, respectively. Periodic
boundary conditions are implemented on the left and right boundaries.
From below and above, the domain is bounded by rigid horizontal walls
made up from monodisperse particles, with the properties as specified
above. The role of the top wall is essentially to contain those few
particles which would be ejected during particularly violent impacts.
However, the upper boundary is positioned sufficiently high that
collisions with this wall are very rare.

%The speed of sound, $c$, in the system will be needed below in order
%to put the results in perspective. We estimate this property by the
%time needed for information to propagate across the domain.
%Specifically, we apply a point force at the top of the granular bed
%and measure the time needed for the force information to reach the
%bottom. When using the kinetic friction model, we find (in
%dimensionless units) $c \approx 2$, while for the static friction
%model, we find slightly larger $c \approx 2.4$. In our simulations,
%we concentrate on the subsonic regime, and consider intruder speeds up
%to $1$ in our dimensionless units.
 
The simulations are typically carried out using $6000$ particles, with
the size of the domain, in units of the mean particle diameter, given
by $L=100$ in the horizontal direction, and the initial height of the
granular bed, given by $W=60$, see Fig.~\ref{fig:setup}. After
settling, the particles form a system of height $\sim 56 d$ for random
polydisperse systems.

The volume fraction, $\rho$, occupied by the grains is difficult to
compute precisely due to the presence of a rough (on the grain scale)
upper surface. Furthermore, some variations of $\rho$ may also result
due to different initial configurations. These variations are less
than about $0.01$, with typical $\rho$'s being in the range $0.85 -
0.86$ for the random polydisperse systems. The influence of the change
of simulation parameters, such as polydispersity, force constant, or
gravity, leads to modifications of $\rho$ on the same scale as
different initial conditions. The influence of different initial
conditions on large-scale features of the results, such as the final
penetration depth, is minor, and therefore we expect that the
influence of slight variations of $\rho$ reported above is not
significant.

\section{Conclusions}

The preceding discussion is a summary of the grain-scale dynamics
observed in a series of studies consisting of two-dimensional granular
impact experiments and simulations. We have approached this problem
from the perspective of a macroscopic force law which is dominated by
a velocity-squared drag force. By using photoelastic disks and a
high-speed camera, we have experimental access to the trajectory of
the intruder as well as the force dynamics of the granular material at
the smallest relevant space and time scales in the system. This yields
microscopic data which has been previously inaccessible
experimentally.{\bf The simulations parallel the experimental results
  in the applicable parameter range.} The results from this work can
be summarized as follows:

\begin{enumerate}

\item[1)] This force law, which has typically been used for impacts
  into three-dimensional granular systems composed of sand or glass
  beads, is also effective in describing impacts into two-dimensional
  systems of photoelastic disks. These disks have a range of
  stiffnesses that range from much softer than--to comparable to--sand
  and glass beads. The frictional properties of the 2D particles are
  comparable to those of sand or glass beads. Thus, many of the
  microscopic insights gained in the two-dimensional system should
  also be applicable to impacts in three-dimensional systems.

\item[2)] Even though this macroscopic force law is effective at
  capturing the trajectories on average, the instantaneous force is
  highly fluctuating in space and time. Physically, these fluctuations
  correspond to intermittent acoustic activity which occurs at the
  leading edge of the intruder as it moves. Thus, at the grain-scale,
  the intruder's energy is transferred to the granular material in a
  series of intermittent collisions with force-chain-like structures.

\item[3)] The deceleration is dominated by a velocity-squared drag
  term, and the strength of this collisional term, $h(z)$, shows
  significant dependence on intruder shape, even when holding mass and
  cross-sectional width constant. This effect can be understood with a
  collisional model which assumes that the origin of the drag force on
  the intruder is a series of random collisions with stationary
  clusters of grains. This model is extremely effective at capturing
  the drag coefficient over a wide range of intruder shapes, as well
  as capturing the dynamics of off-axis rotations. It is noteworthy
  that `collisions' between the intruder and the granular bed, as as
  observed directly in the photoelastic images, are typically normal
  to the intruder boundary. This effect is incorporated in the model
  by explicitly assuming that friction between the grains and the
  leading edge of the intruder can be ignored.  The success of the
  model in quantitatively characterizing the drag forces on an
  intruder supports this assumption.

\item[4)] When the rate of force transmission is comparable to the
  intruder speed, the nature of the granular force response changes,
  and the Poncelet model breaks down. The strong force network (force
  chains) does not decay as the the intruder advances, but rather,
  continues to grow at the front of the shock. Contacts in the
  compressed region have relatively large deformations. Since the
  particles are incompressible, these large deformations lead to the
  formation of additional contacts which further strenthen the
  compressed part of the sample which exists between the leading edge
  of the intruder and the shock front.

\end{enumerate}

One obvious direction for future research would be to apply the
collisional model to impacts in three dimensions. A number of 3D
experiments with comparable impact and granular force propagation
speeds also find $v^2$ scaling for the dynamic part of the drag. An
extension of the model to 3D is straight forward, and would involve
extensions of Eqs.~\ref{eqn:collforce}~\ref{eqn:torque-tot2} to the
leading surface of a 3D intruder. The fact that stresses from
collisions can be determined within $O(1)$ normalization suggests that
it may be possible to use such an approach for applications, such as
maximizing or minimizing the inertial drag or understanding the
stability and dynamics of rotations of granular intruders as a
function of intruder shape.

\bibliography{References}

\end{document}